%% file: main.tex
\documentclass[10pt,twocolumn,letterpaper]{article}

\usepackage[pagenumbers]{cvpr} 

\input{preamble}

\input{math_commands.tex}

\usepackage[accsupp]{axessibility}
\usepackage[symbol]{footmisc}
\usepackage[pagebackref,breaklinks,colorlinks,citecolor=cvprblue]{hyperref}
\usepackage{pifont}
\usepackage[ruled,vlined]{algorithm2e}
\usepackage{listings}

\definecolor{cvprblue}{rgb}{0.21,0.49,0.74}
\lstdefinestyle{overleaf}{
    backgroundcolor=\color[rgb]{0.95,0.95,0.92},   
    commentstyle=\color[rgb]{0,0.6,0},
    keywordstyle=\color{magenta},
    numberstyle=\tiny\color[rgb]{0.5,0.5,0.5},
    stringstyle=\color[rgb]{0.58,0,0.82},
    basicstyle=\ttfamily\footnotesize,
    breakatwhitespace=false,         
    breaklines=true,                 
    captionpos=b,                    
    keepspaces=true,                 
    numbers=left,                    
    numbersep=5pt,                  
    showspaces=false,                
    showstringspaces=false,
    showtabs=false,                  
    tabsize=2
}
\lstdefinestyle{mocov3}{
  backgroundcolor=\color{white},
  basicstyle=\fontsize{7.5pt}{7.5pt}\ttfamily\selectfont,
  columns=fullflexible,
  breaklines=true,
  captionpos=b,
  emph={self, with},
  emphstyle=\fontsize{7.5pt}{7.5pt}\color[rgb]{0.85,0.18,0.50},
  commentstyle=\fontsize{7.5pt}{7.5pt}\color[rgb]{0.25,0.5,0.5},
  keywordstyle=\fontsize{7.5pt}{7.5pt}\color[rgb]{0.85,0.18,0.50},
}
\newcommand{\methodname}{{QN-Mixer}\xspace} 
\newcommand{\regterm}{{Incept-Mixer}\xspace}
\definecolor{trolleygrey}{rgb}{0.5, 0.5, 0.5}
\definecolor{trolleygrey}{RGB}{128, 128, 128}
\definecolor{modif2}{HTML}{FF7F00}
\definecolor{softblue}{RGB}{0, 68, 170}
\definecolor{softred}{RGB}{255, 85, 85}
\definecolor{softgreen}{RGB}{44, 160, 44}
\definecolor{softgray}{RGB}{200, 200, 200}
\newcommand{\seg}[1]{\noindent \textbf{#1}~}
\newcommand{\cmark}{\text{\ding{51}}}

\newcommand{\grayxmark}{\textcolor{softgray}{\ding{55}}}
\newcommand{\customfootnotetext}[2]{{
  \renewcommand{\thefootnote}{#1}
  \footnotetext[0]{#2}}}
\def\paperID{15740} 
\def\confName{CVPR}
\def\confYear{2024}

\begin{document}
\title{QN-Mixer: A \underline{Q}uasi-\underline{N}ewton MLP-\underline{Mixer} Model \\ for Sparse-View CT Reconstruction}

\author{
Ishak Ayad\textsuperscript{1,2*$\dagger$} \qquad
Nicolas Larue\textsuperscript{1,3$\dagger$} \qquad
Ma\"{\i} K. Nguyen$^{1}$ \\
$^{1}$ETIS (UMR 8051), CY Cergy Paris University, ENSEA, CNRS, France\\
$^{2}$AGM (UMR 8088), CY Cergy Paris University, CNRS, France\\
$^{3}$University of Ljubljana, Slovenia \\
{\tt\small ishak.ayad@cyu.fr}
}

\twocolumn[{
      \maketitle
      \vspace{-2em}
      \includegraphics[width=\linewidth]{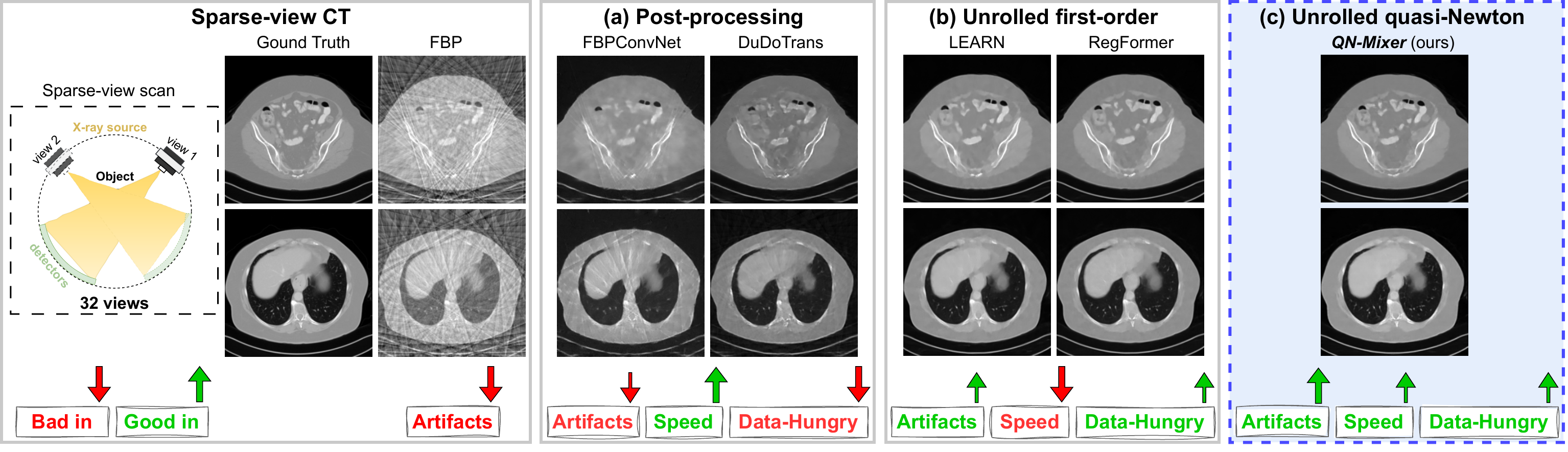}
      \captionof{figure}{\textbf{CT Reconstruction with 32 views of State-of-the-Art Methods}. Comparative analysis with post-processing and first-order unrolling networks highlights \methodname's superiority in artifact removal, training time, and data efficiency.}
      \vspace{1em}
      \label{fig:teaser}
    }]

\customfootnotetext{*}{Corresponding author. ${\dagger}$ Equal contribution.}

\input{sec/0_abstract}
\input{sec/1_intro}

\input{sec/2_related_works}

\input{sec/3_methodology}

\input{sec/4_experiments}
\input{sec/5_conclusion}

\section{Acknowledgments}
This work was granted access to the HPC resources of IDRIS under the allocation 2021-[AD011012741] provided by GENCI and supported by DIM Math Innov funding.

{
  \small
  \bibliographystyle{ieeenat_fullname}
  \bibliography{main}
}

\input{sec/X_suppl}

\end{document}

%% file: preamble.tex
\usepackage[dvipsnames]{xcolor}

\newcommand{\subtabtitle}[2]{
    \multicolumn{#2}{l}{{\small\textcolor{trolleygrey}{\textit{#1}}}}
}

\usepackage{lipsum}
\newcommand{\crefcustom}[2]{\hyperref[#1]{\color{black}\text{#2}}}

\usepackage{wrapfig}

\usepackage{arydshln} 

\usepackage{titletoc}

\usepackage{multirow}

%% file: math_commands.tex
\usepackage{amsmath,amsfonts,bm}

\newcommand{\regfunc}[1]{{\mathcal #1}}

\def\eqref#1{equation~\ref{#1}}

\def\1{\bm{1}}

\def\vd{{\bm{d}}}
\def\ve{{\bm{e}}}
\def\vf{{\bm{f}}}

\def\vr{{\bm{r}}}
\def\vs{{\bm{s}}}

\def\vx{{\bm{x}}}
\def\vy{{\bm{y}}}
\def\vz{{\bm{z}}}

\def\mA{{\bm{A}}}
\def\mB{{\bm{B}}}

\def\mH{{\bm{H}}}
\def\mI{{\bm{I}}}

\def\mW{{\bm{W}}}

\DeclareMathAlphabet{\mathsfit}{\encodingdefault}{\sfdefault}{m}{sl}
\SetMathAlphabet{\mathsfit}{bold}{\encodingdefault}{\sfdefault}{bx}{n}

\def\sN{{\mathbb{N}}}

\newcommand{\R}{\mathbb{R}}

\newcommand{\normltwo}{L^2}

\newcommand{\argmin}[1]{\underset{#1}{\operatorname{arg}\,\operatorname{min}}\;}

\newcommand\norm[1]{\left\lVert#1\right\rVert}

%% file: sec/0_abstract.tex
\begin{abstract}
Inverse problems span across diverse fields. In medical contexts, computed tomography (CT) plays a crucial role in reconstructing a patient's internal structure, presenting challenges due to artifacts caused by inherently ill-posed inverse problems.
Previous research advanced image quality via post-processing and deep unrolling algorithms but faces challenges, such as extended convergence times with ultra-sparse data. Despite enhancements, resulting images often show significant artifacts, limiting their effectiveness for real-world diagnostic applications.
We aim to explore deep second-order unrolling algorithms for solving imaging inverse problems, emphasizing their faster convergence and lower time complexity compared to common first-order methods like gradient descent.
In this paper, we introduce \methodname, an algorithm based on the quasi-Newton approach. We use learned parameters through the BFGS algorithm and introduce \regterm, an efficient neural architecture that serves as a non-local regularization term, capturing long-range dependencies within images.
To address the computational demands typically associated with quasi-Newton algorithms that require full Hessian matrix computations, we present a memory-efficient alternative. Our approach intelligently downsamples gradient information, significantly reducing computational requirements while maintaining performance.
The approach is validated through experiments on the sparse-view CT problem, involving various datasets and scanning protocols, and is compared with post-processing and deep unrolling state-of-the-art approaches.
Our method outperforms existing approaches and achieves state-of-the-art performance in terms of SSIM and PSNR, all while reducing the number of unrolling iterations required.
\end{abstract}

%% file: sec/1_intro.tex
\section{Introduction}
\label{sec:intro}

Computed tomography (CT) is a widely used imaging modality in medical diagnosis and treatment planning, delivering intricate anatomical details of the human body with precision. Despite its success, CT is associated with high radiation doses, which can increase the risk of cancer induction~\cite{ct_outlook}. Adhering to the ALARA principle (As Low As Reasonably Achievable)~\cite{alara}, the medical community emphasizes minimizing radiation exposure to the lowest level necessary for accurate diagnosis. Numerous approaches have been proposed to reduce radiation doses while maintaining image quality. Among these, sparse-view CT emerges as a promising solution, effectively lowering radiation doses by subsampling the projection data, often referred to as the sinogram. Nonetheless, reconstructed images using the well-known Filtered Back Projection (FBP) algorithm~\cite{fbp}, suffer from pronounced streaking artifacts (see \cref{fig:teaser}), which can lead to misdiagnosis. The challenge of effectively reconstructing high-quality CT images from sparse-view data is gaining increasing attention in both the computer vision and medical imaging communities.

With the success of deep learning spanning diverse domains, initial image-domain techniques~\cite{redcnn, fbpconvnet, ddnet, leemwcnn, gloredi} have been introduced as post-processing tasks on the FBP reconstructed images, exhibiting notable accomplishments in artifact removal and structure preservation. However, the inherent limitations of these methods arise from their constrained receptive fields, leading to challenges in effectively capturing global information and, consequently, sub-optimal results. 

To address this limitation, recent advances have seen a shift toward a dual-domain approach~\cite{dudonet, hdnet, dudotrans, ddptransformer}, where post-processing methods turn to the sinogram domain. In this dual-domain paradigm, deep neural networks are employed to perform interpolation tasks on the sinogram data~\cite{ssnet, sinogramcom}, facilitating more accurate image reconstruction. Despite the significant achievements of post-processing and dual-domain methods, they confront issues of interpretability and performance limitations, especially when working with small datasets and ultra-sparse-view data, as shown in \cref{fig:teaser}. 
To tackle these challenges, deep unrolling networks have been introduced~\cite{learnedpd, cnngrad, learn, admmct, cheng2020, dior, fistanet, regformer}.
Unrolling networks treat the sparse-view CT reconstruction problem as an optimization task, resulting in a first-order iterative algorithm like gradient descent, which is subsequently unrolled into a deep recurrent neural network in order to learn the optimization parameters and the regularization term. Like post-processing techniques, unrolling networks have been extended to the sinogram domain~\cite{drone, learnpp} to perform interpolation task. 

Unrolling networks, as referenced in~\cite{endtoend, humusnet, adafactor}, exhibit remarkable performance across diverse domains. However, they suffer from slow convergence and high computational costs, as illustrated in \cref{fig:teaser}, necessitating the development of more efficient alternatives~\cite{optimus}.
More specifically, they confront two main issues: 
\emph{Firstly}, they frequently grapple with capturing long-range dependencies due to their dependence on locally-focused regularization terms using CNNs. This limitation results in suboptimal outcomes, particularly evident in tasks such as image reconstruction.
\emph{Secondly}, the escalating computational costs of unrolling methods align with the general trend of increased complexity in modern neural networks. This escalation not only amplifies the required number of iterations due to the algorithm's iterative nature but also contributes to their high computational demand.

To tackle the aforementioned issues, we introduce a novel second-order unrolling network for sparse-view CT reconstruction.
\emph{In particular}, to enable the learnable regularization term to apprehend long-range interactions within the image, we propose a non-local regularization block termed \textbf{\regterm}. Drawing inspiration from the multi-layer perceptron mixer~\cite{mlpmixer} and the inception architecture~\cite{inception}, it is created to combine the best features from both sides: capturing long-range interactions from the attention-like mechanism of MLP-Mixer and extracting local invariant features from the inception block. This block facilitates a more precise image reconstruction.
\emph{Second}, to cut down on the computational costs associated with unrolling networks, we propose to decrease the required iterations for convergence by employing second-order optimization methods such as~\cite{qn, saddleqn}. We introduce a novel unrolling framework named \textbf{\methodname}. 
Our approach is based on the quasi-Newton method that approximate the Hessian matrix using the Broyden-Fletcher-Goldfarb-Shanno (BFGS) update~\cite{bfgs, qnbfgs, ctbfgs}.
Furthermore, we reduce memory usage by working on a projected gradient (latent gradient), preserving performance while reducing the computational cost tied to Hessian matrix approximation. This adaptation enables the construction of a deep unrolling network, showcasing superlinear convergence.
\noindent Our contributions are summarized as follows:
\begin{itemize}
    \item We introduce a novel second-order unrolling network coined \textbf{\methodname} where the Hessian matrix is approximated using a latent BFGS algorithm with a deep-net learned regularization term.
    \newline
    \item We propose \textbf{\regterm}, a neural architecture acting as a non-local regularization term. \regterm integrates deep features from inception blocks with MLP-Mixer, enhancing multi-scale information usage and capturing long-range dependencies.
    \newline
    \item We demonstrate the effectiveness of our proposed method when applied to the sparse-view CT reconstruction problem on an extensive set of experiments and datasets.
    We show that our method outperforms state-of-the-art methods in terms of quantitative metrics while requiring less iterations than first-order unrolling networks.
\end{itemize}

%% file: sec/2_related_works.tex
\section{Related Works}
\label{sec:related}

In this section, we present prior work closely related to our paper. We begin by discussing the general framework for unrolling networks in \cref{sec:background}, which is based on the gradient descent algorithm. Subsequently, in \cref{sec:related_postprocess} and \cref{sec:related_unrolled}, we delve into state-of-the-art methods in post-processing and unrolling networks, respectively.

\subsection{Background}
\label{sec:background}

\seg{Inverse Problem Formulation for CT.}
Image reconstruction problem in CT can be mathematically formalized as the solution to a linear equation in the form of:
\begin{align}
\vy=\mA\vx,
\label{eq:linear}
\end{align}
\noindent where $\vx \in \R^n$ is the (unknown) object to reconstruct with $n=h \times w$, $\vy \in \R^m$ is the data (i.e. sinogram), where $m=n_v \times n_d$, $n_v$ and $n_d$ denote the number of projection views and detectors, respectively. $\mA \in \R^{n \times m}$ is the forward model (i.e. discrete Radon transform~\cite{radon}). The goal of CT image reconstruction is to recover the (unknown) object, $\vx$, from the observed data $\vy$. As the problem is ill-posed due to the missing data, the linear system in \cref{eq:linear} becomes underdetermined and may have infinite solutions. Hence, reconstructed images suffer from artifacts, blurring, and noise. To address this issue, iterative reconstruction algorithms are utilized to minimize a regularized objective function with a $\normltwo$ norm constraint:
\begin{align}
        \hat{\vx}=\argmin{\vx} J(\vx)=\frac{\lambda}{2} \norm{\mA\vx - \vy}^2_2 + \regfunc{R}(\vx),
        \label{eq:objective}
\end{align}
\noindent where $\regfunc{R}(\vx)$ is the regularization term, balanced with the weight $\lambda$.
Those ill-posed problems were initially addressed using optimization techniques, such as the truncated singular value decomposition (SVD) algorithm~\cite{svd}, or iterative approaches like the algebraic reconstruction technique (ART)~\cite{art}, simultaneous ART (SART)~\cite{sart}, conjugate gradient for least squares (CGLS)~\cite{cgls}, and total generalized variation regularization (TGV)~\cite{tgv}. Additionally, techniques such as total variation~\cite{tv} and Tikhonov regularization~\cite{tikhonov} can be employed to enhance reconstruction results.

\seg{Deep Unrolling Networks.}
By assuming that the regularization term in \cref{eq:objective} (i.e. $\regfunc{R}$) is differentiable and convex, a simple gradient descent scheme can be applied to solve the optimization problem:
\begin{align}
\begin{aligned}
    &\vx_{t+1}=\vx_{t} - \alpha \nabla_\vx J(\vx_t), \\
    &\text{where} \quad \nabla_\vx J(\vx_t)=\lambda \mA^{\dagger}\left(\mA\vx_t - \vy\right) + \nabla_\vx \regfunc{R}(\vx_t).
\end{aligned}    
\label{eq:gradient_solution}
\end{align}
\noindent Here, $\alpha$ represents the step size (i.e. search step), and $\mA^{\dagger}$ is the pseudo-inverse of $\mA$. 

Previous research~\cite{wu2017iterative, cnngrad} has emphasized the limitations of optimization algorithms, such as the manual selection of the regularization term and the optimization hyper-parameters, which can negatively impact their performance, limiting their clinical application. Recent advancements in deep learning techniques have enabled automated parameter selection directly from the data, as demonstrated in~\cite{lunz2018, learn, learnpp, dior, kobler2020total, mukherjee2021endtoend}. By allowing the terms in \cref{eq:gradient_solution} to be dependent on the iteration, the gradient descent iteration becomes:
\begin{align}
    \vx_{t+1}=\vx_{t} - \lambda_{t} \mA^{\dagger}\left(\mA\vx_{t} - \vy\right) + \regfunc{G}(\vx_{t}),
\label{eq:unrolled_solution}
\end{align}
\noindent where $\regfunc{G}$ is a learned mapping representing the gradient of the regularization term. It is worth noting that the step size $\alpha$ in \cref{eq:gradient_solution} is omitted as it is redundant when considering the learned components of the regularization term. Finally, \cref{eq:unrolled_solution} is unrolled into a deep recurrent neural network in order to learn the optimization parameters.

\subsection{Post-processing Methods}
\label{sec:related_postprocess}
\begin{figure*}[t] 
    \centering 
    \includegraphics[width=1.0\textwidth]{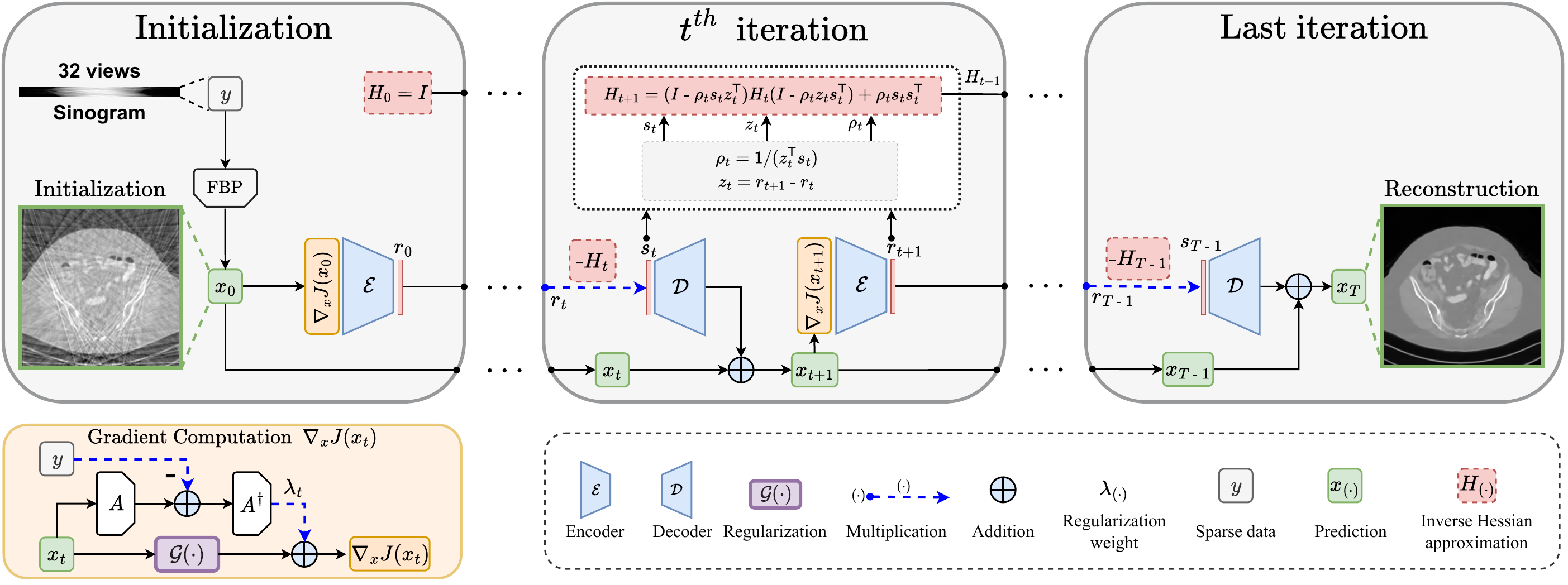}
    \caption{\textbf{Overall structure of the proposed \methodname} for sparse-view CT reconstruction, unrolled from \cref{algo:qnmixer}. The method leverages the advantages of the quasi-Newton method for faster convergence while incorporating a latent BFGS update.}
    \label{fig:flow} 
\end{figure*}
Recent advances in sparse-view CT reconstruction leverage two main categories of deep learning methods: post-processing and dual-domain approaches.
Post-processing methods, including RedCNN~\cite{redcnn}, FBPConvNet~\cite{fbpconvnet}, and DDNet~\cite{ddnet}, treat sparse-view reconstruction as a denoising step using FBP reconstructions as input. While effective in addressing artifacts and reducing noise, they often struggle with recovering global information from extremely sparse data.
To overcome this limitation, dual-domain methods integrate sinograms into neural networks for an interpolation task, recovering missing data~\cite{ssnet, sinogramcom}. Dual-domain methods, surpassing post-processing ones, combine information from both domains.
DuDoNet~\cite{dudonet}, an initial dual-domain method, connects image and sinogram domains through a Radon inversion layer. Recent Transformer-based dual-domain methods, such as DuDoTrans~\cite{dudotrans} and DDPTransformer~\cite{ddptransformer}, aim to capture long-range dependencies in the sinogram domain, demonstrating superior performance to CNN-based methods.

\seg{Self-supervised learning.}
SSL methods~\cite{n2inv, ei, IntraTomo, N2S, SACNN}, have been applied for CT reconstruction.
For instance,~\cite{ei} proposed an equivariant imaging paradigm through a training strategy that enforces measurement consistency and equivariance conditions.
To ensure equitable comparisons, we focus on supervised methods in this work.

\subsection{Advancements in Deep Unrolling Networks}
\label{sec:related_unrolled}
Unrolling networks constitute a line of work inspired by popular optimization algorithms used to solve \cref{eq:objective}. Leveraging the iterative nature of optimization algorithms, as presented in \cref{eq:unrolled_solution}, unrolling networks aim to directly learn optimization parameters from data. These methods have found success in various inverse problems, including sparse-view CT~\cite{learn, fistanet, drone, learnpp, regformer}, limited-angle CT~\cite{dior, admmct, cheng2020}, low-dose CT~\cite{learnedpd, cnngrad}, and compressed sensing MRI~\cite{endtoend, humusnet}.

\seg{First-order.}
One pioneering unrolling network, Learned Primal-Dual reconstruction~\cite{learnedpd}, replaces traditional proximal operators with CNNs. 
In contrast, LEARN~\cite{learn} and LEARN++~\cite{learnpp} directly unroll the optimization algorithm from \cref{eq:unrolled_solution} into a deep recurrent neural network. 
More recently, Transformers~\cite{attention, swint} have been introduced into unrolling networks, such as RegFormer~\cite{regformer} and HUMUS-Net~\cite{humusnet}. While achieving commendable performance, these methods require more computational resources than traditional CNN-based unrolling networks and incur a significant memory footprint due to linear scaling with the number of unrolling iterations.

\seg{Second-order.}
To address this, a new category of unrolling optimization methods has emerged~\cite{optimus}, leveraging second-order techniques like the quasi-Newton method~\cite{qn, bfgs, qnbfgs}. 
These methods converge faster, reducing computational demands, but struggle with increased memory usage due to Hessian matrix approximation and their application is limited to small-scale problems~\cite{saddleqn, ctbfgs}. 
In contrast our method propose a memory-efficient approach by operating within the latent space of gradient information (i.e. $\nabla_\vx J(\vx)$ in \cref{eq:gradient_solution}).

%% file: sec/3_methodology.tex
\begin{algorithm}[b]
    \caption{Quasi-Newton for sparse-view CT}
    \KwData{$\vy$ (sparse sinogram)}
    Manual choice of the regularization term $\regfunc{R}$\;
    $\mH_0 \gets \mI^{n \times n}$\;
    $\vx_0 \gets \mA^{\dagger}\vy$\;
    \For{\texttt{$t \in \{0, \dots, T-1\}$}}{               
        $\vs_{t} \gets -\mH_t \nabla_{\vx} J(\vx_t)$
        
        $\vx_{t+1} \gets \vx_t + \vs_t$

        $\vz_t \gets \nabla_{\vx} J(\vx_{t+1}) - \nabla_{\vx} J(\vx_t)$
              
        $\rho_t \gets 1/(\vz_t^\mathsf{T} \vs_t)$

        $\mH_{t+1} \gets (\mI - \rho_t \vs_t \vz_t^\mathsf{T}) \mH_t (\mI - \rho_t \vz_t \vs_t^\mathsf{T}) + \rho_t \vs_t \vs_t^\mathsf{T}$
    }
    \label{algo:ctqn}
\end{algorithm}

\section{Methodology}
\label{sec:method}

\methodname is a novel second-order unrolling network inspired by the quasi-Newton (\cref{sec:method_qn}) method. It approximates the inverse Hessian matrix with a latent BFGS algorithm and includes a non-local regularization term, \regterm, designed to capture non-local relationships (\cref{sec:incept_mixer}). To cope with the significant computational burden associated with the full approximation of the inverse Hessian matrix, we use a latent BFGS algorithm (\cref{sec:latent_bfgs}). An overview of the proposed method is depicted in \cref{fig:flow}, and the complete algorithm is presented in \cref{sec:qnmixer_algo}.

\subsection{Quasi-Newton method}
\label{sec:method_qn}

The quasi-Newton method can be applied to solve \cref{eq:objective} and the iterative optimization solution is expressed as:
\begin{align}
    \vx_{t+1}=\vx_{t} - \alpha_t \mH_t \nabla_\vx J(\vx_t), 
    \label{eq:qn_solution}
\end{align}
\noindent where $\mH_t \in \R^{n \times n}$ represents the inverse Hessian matrix approximation at iteration $t$, and $\alpha_t$ is the step size. The BFGS method updates the Hessian matrix approximation in each iteration. This matrix is crucial for understanding the curvature of the objective function around the current point, guiding us to take more efficient steps and avoiding unnecessary zigzagging.
In the classical BFGS approach, the line search adheres to Wolfe conditions~\cite{bfgs, qnbfgs}. A step size of $\alpha_t=1$ is attempted first, ensuring eventual acceptance for superlinear convergence~\cite{qn}. 
In our approach, we adopt a fixed step size of $\alpha_t=1$.
The algorithm is illustrated in \cref{algo:ctqn}.

\subsection{Regularization term: \regterm}
\label{sec:incept_mixer}

\begin{figure}
  \centering
   \includegraphics[width=\linewidth]{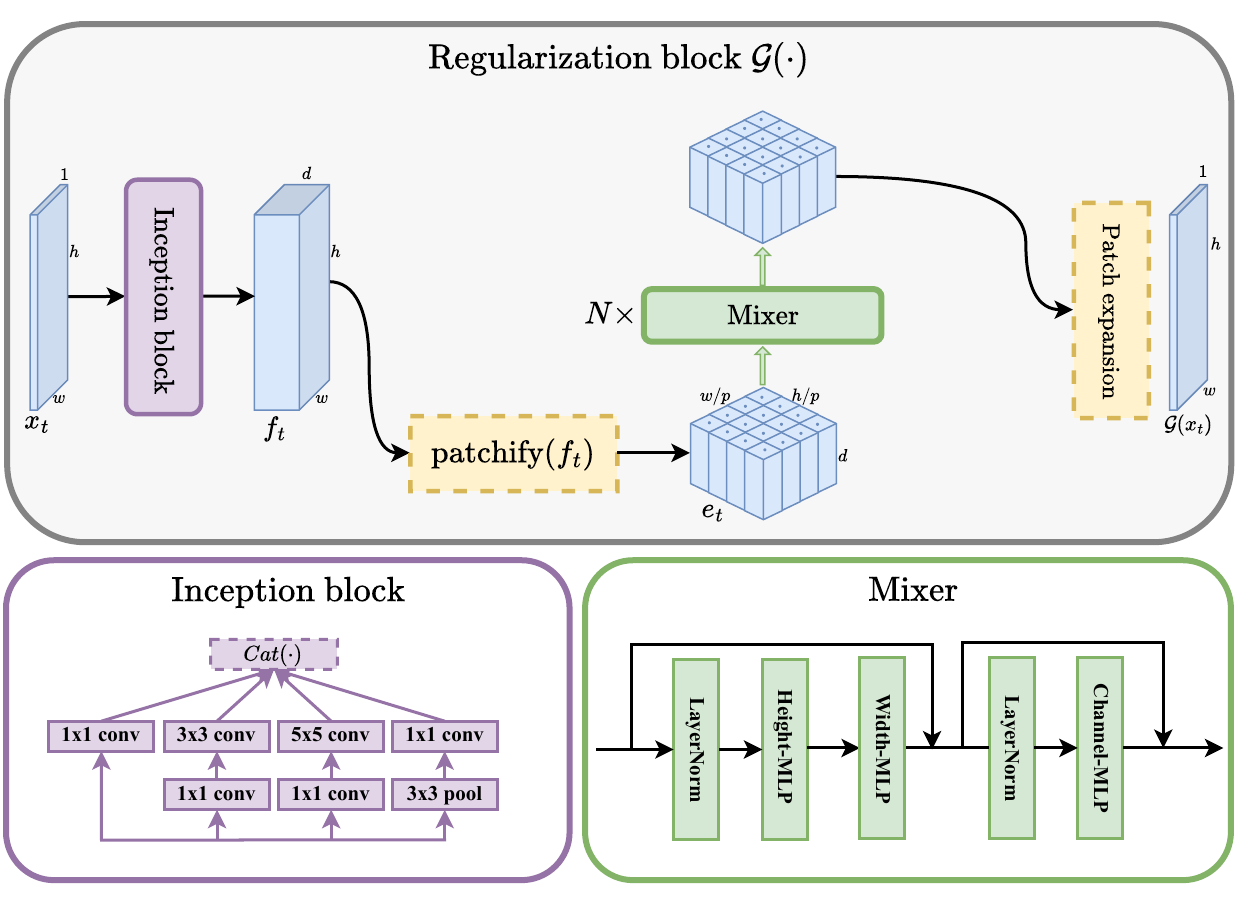}
   \caption{\textbf{Architecture of our regularization block}. It is referred to as ``\textbf{\regterm}" and denoted as $\regfunc{G}$ in \cref{eq:unrolled_solution}}
   \label{fig:inceptmixer}
   \vspace{-1.5em}
\end{figure}

Recent research on unrolling networks has often focused on selecting the representation of the regularization term gradient (i.e. $\regfunc{G}$ in \cref{eq:unrolled_solution}), ranging from conv-nets~\cite{learn, learnpp, endtoend} to more recent attention-based nets~\cite{regformer, humusnet}. In alignment with this trend, we introduce a non-local regularization block named \textbf{\regterm} and depicted in, \cref{fig:inceptmixer}. This block is crafted by drawing inspiration from both the multi-layer perceptron mixer~\cite{mlpmixer} and the inception architecture~\cite{inception}, leveraging the strengths of each: capturing long-range interactions through the attention-like mechanism of MLP-Mixer and extracting local invariant features from the inception block. This design choice is evident in the ablation study (see \cref{tab:reg_ablation}) where \regterm outperforms both alternatives.

Starting from an image $\vx_t \in \R^{h \times w \times c}$ at iteration $t$, we pass it through an Inception block to create a feature map $\vf_t \in \R^{h \times w \times d}$, where $d$ is the depth of features. Subsequently, $\vf_t$ undergoes patchification using a CNN with a kernel size and stride of $p$, representing the patch size. This process yields patch embeddings, $\ve_t=\mathrm{patchify}(\vf_t) \in \R^{\frac{h}{p} \times \frac{w}{p} \times d}$. These embeddings are then processed through a \textbf{Mixer Layer} with token and channel MLPs, layer normalization, and skip connections for inter-layer information flow, following~\cite{mlpmixer}:
\begin{equation}
\mathrm{Mixer}(\ve_t) =\mathrm{Mix}(\mathrm{MLP}_{\texttt{c}},
\mathrm{Mix}(\left[ \mathrm{MLP}_{\texttt{h}}, \mathrm{MLP}_{\texttt{w}} \right], \ve_t),
\end{equation}
where $\mathrm{Mix}(\mathrm{Layer}, \ve_t)=\mathrm{Layer}(\mathrm{LN}(\ve_t)) + \ve_t$, with $\mathrm{LN}$ as layer normalization. $\mathrm{MLP}_{\texttt{h}}$, $\mathrm{MLP}_{\texttt{w}}$ are applied to height and width features, respectively, and $\mathrm{MLP}_{\texttt{c}}$ to rows and shared.
Finally, after $N$ such mixer layers, the regularized sample is transformed back to an image through a patch expansion step to obtain $\regfunc{G}(\vx_{t})$. Consequently, the iterative optimization solution is as follows:
\begin{align}
\begin{aligned}
    &\vx_{t+1} = \vx_{t} - \mH_t \nabla_\vx J(\vx_t), \\
    &\text{where} \quad \nabla_\vx J(\vx_t)=\lambda_t \mA^{\dagger}\left(\mA\vx_t - \vy\right) + \regfunc{G}(\vx_{t}).
\end{aligned}    
\label{eq:qn_unrolled_solution}
\end{align}
\noindent Here, $\regfunc{G}(\vx_{t})$ denotes the \regterm model, representing the learned gradient of the regularization term. 

\subsection{Latent BFGS update}
\label{sec:latent_bfgs}
We propose a memory-efficient latent BFGS update.
Drawing inspiration from LDMs~\cite{ldms},
at step $t$, given the gradient value $\nabla_{\vx} J(\vx_t) \in \R^{h \times w \times c}$, the encoder $\mathcal{E}$ encodes it into a latent representation $\vr_t = \mathcal{E}(\nabla_{\vx} J(\vx_t)) \in \R^{l_h \cdot l_w}$. Importantly, the encoder downsamples the gradient by a factor $\vf_{\mathcal{E}}=\frac{h}{h_l}=\frac{w}{w_l}$. Throughout the paper, we explore different downsampling factors (see \cref{tab:hess_size}) $\vf_{\mathcal{E}}=2^{k}$, where $k \in \sN$ is the number of downsampling stacks. Encoding the gradient reduces the optimization variable size of BFGS (i.e. $\mH_t \in \R^{(l_h \cdot l_w) \times (l_h \cdot l_w)}$), thereby decreasing the computational cost associated with high memory demand. The direction is then computed in the latent space $s_t=-\mH_t \vr_t$, and finally, the decoder $\mathcal{D}$ reconstructs the update from the latent direction, giving $\mathcal{D}(s_t)=\mathcal{D}(-\mH_t \mathcal{E}(\nabla_{\vx} J(\vx_t))) \in \R^{h \times w \times c}$. It is noteworthy that $\mathcal{E}$ and $\mathcal{D}$ are shared across the algorithm iterations, as shown in \cref{fig:flow}.

\subsection{Proposed algorithm of \methodname}
\label{sec:qnmixer_algo}

\begin{algorithm}[htpb]
    \caption{\methodname (latent BFGS update)}
    \KwData{$\vy$ (sparse sinogram)}
    $\mH_0 \gets \mI^{(l_h \cdot l_w) \times (l_h \cdot l_w)}$\;
    $\vx_0 \gets \mA^{\dagger}\vy$\;
    $\vr_{0} \gets \mathcal{E}(\nabla_{\vx} J(\vx_0))$\;
    \For{\texttt{$t \in \{0, \dots, T-1\}$}}{        
        $\vs_{t} \gets -\mH_t \vr_t$
        
        $\vx_{t+1} \gets \vx_t + \mathcal{D}(\vs_t)$
        
        $\vr_{t+1} \gets \mathcal{E}(\nabla_{\vx} J(\vx_{t+1}))$
        
        $\vz_t \gets \vr_{t+1} - \vr_t$
              
        $\rho_t \gets 1/(\vz_t^\mathsf{T} \vs_t)$

        $\mH_{t+1} \gets (\mI - \rho_t \vs_t \vz_t^\mathsf{T}) \mH_t (\mI - \rho_t \vz_t \vs_t^\mathsf{T}) + \rho_t \vs_t \vs_t^\mathsf{T}$
    }
    \label{algo:qnmixer}
\end{algorithm}

Our method, builds on the BFGS update~\cite{bfgs, qnbfgs} rank-one approximation for the inverse Hessian. This approximation serves as a preconditioning matrix, guiding the descent direction. In contrast to~\cite{optimus}, which directly learns the inverse Hessian approximation from data, our approach incorporates the mathematical equations of the BFGS algorithm for more accurate approximations.
The full \methodname algorithm is illustrated in \cref{algo:qnmixer}.

%% file: sec/4_experiments.tex
\input{tab/aapm_quantitative}
\begin{figure*}[t] 
    \centering 
    \includegraphics[width=1.0\textwidth]{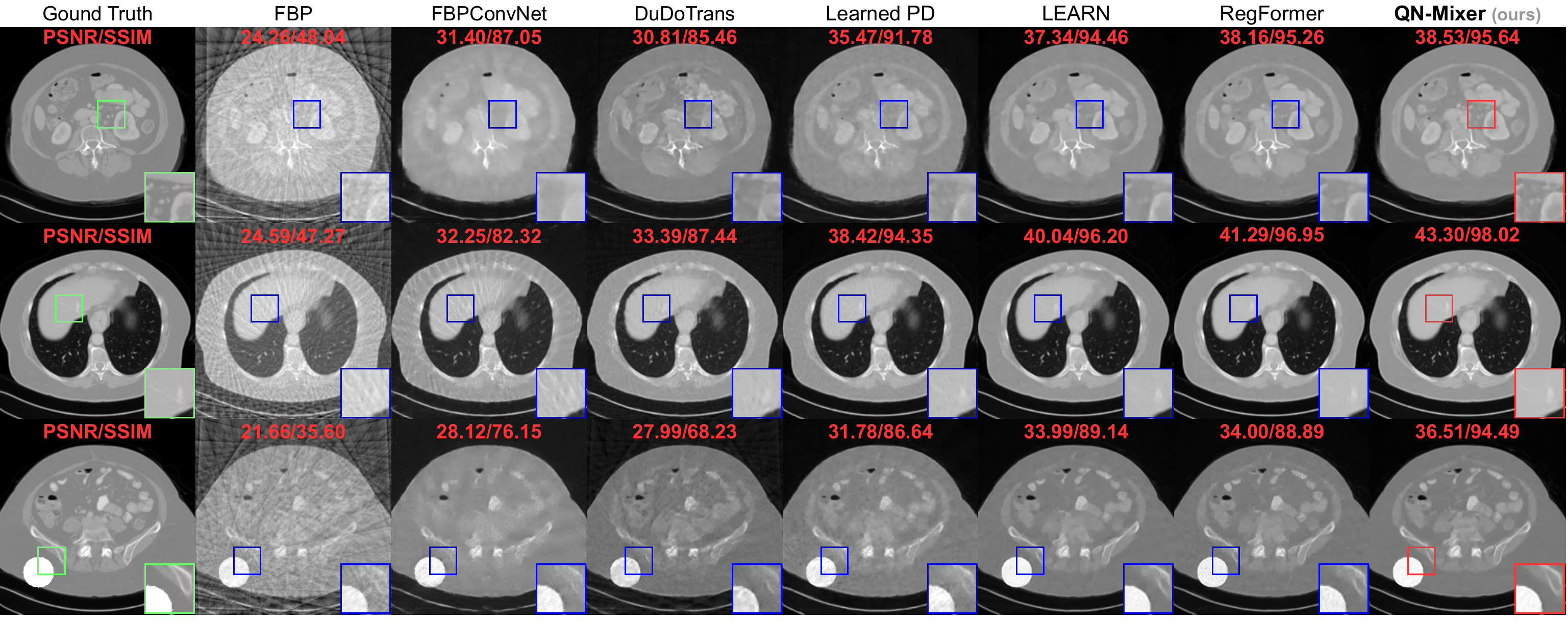}
    \vspace{-0.5cm}
    \caption{\textbf{Visual comparison on AAPM}. From top to bottom: the results under the following conditions: first $(n_v{=}32, N_1)$, second $(n_v{=}64, N_1)$, third $(n_v{=}32, N_0)$. The last row presents out-of-distribution (OOD) results with a randomly overlaid circle on a test image. The display window is set to $\left[-1000, 800\right]$ HU.}    
    \label{fig:aapm_visual} 
    \vspace{-0.1em}
\end{figure*}
\section{Experiments}
\label{sec:exper}
In this section, we initially present our experimental settings, followed by a comparison of our approach with other state-of-the-art CT reconstruction methods. Finally, we delve into the contribution analysis of each component in our model.

\begin{figure*}[t] 
    \centering 
    \includegraphics[width=1.0\textwidth]{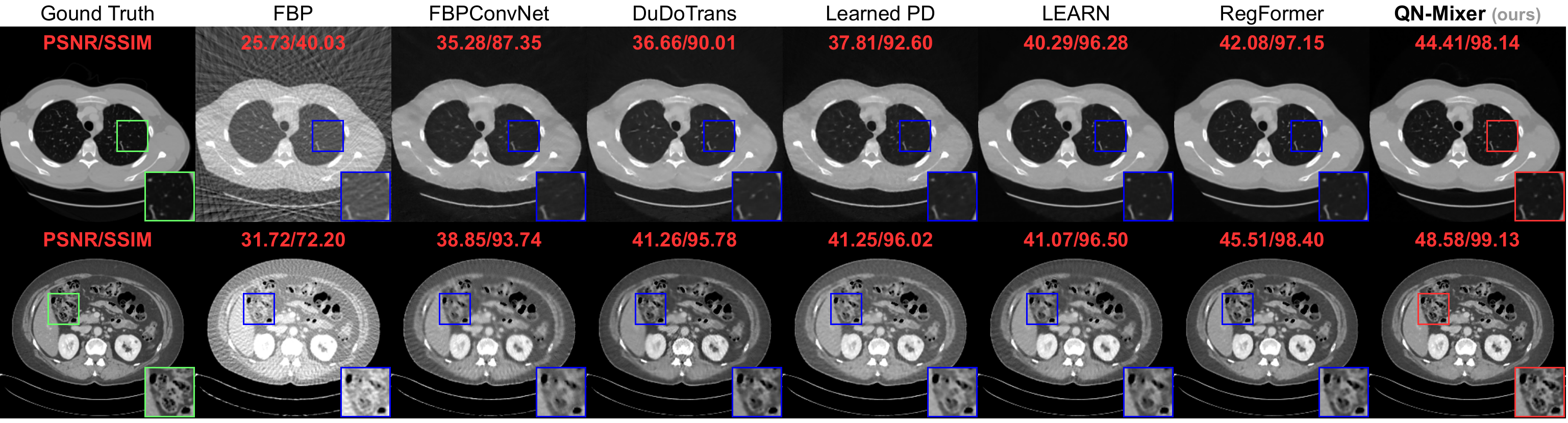}
    \vspace{-0.5cm}
    \caption{\textbf{Visual comparison on DeepLesion} of state-of-the-art methods. Rows display results under different conditions: $(n_v{=}64, N_1)$ and $(n_v{=}128, N_1)$. Display windows are set to $\left[-1000, 800\right]$ HU for the first row and $\left[-200, 300\right]$ HU for the second row.}
    \label{fig:deeplesion_visual} 
\end{figure*}

\subsection{Experimental Setup}
\label{sec:exper_setup}
\seg{Datasets.}
We evaluate our method on two widely used datasets: the ``2016 NIH-AAPM-Mayo Clinic Low-Dose CT Grand Challenge'' dataset (AAPM)~\cite{aapm} and the DeepLesion dataset~\cite{deeplesion}. The AAPM dataset comprises 2378 full-dose CT images from 10 patients, while DeepLesion is the largest publicly accessible multi-lesion real-world CT dataset, including 4427 unique patients.

\seg{Implementation details.} \label{seg:imple_details}
For AAPM, we select $1920$ training images from 8 patients, $244$ validation images from $1$ patient, and $214$ testing images from the last patient. For DeepLesion, we select a subset of $2000$ training images and $300$ testing images randomly from the official splits. All images are resized to $256 \times 256$ pixels. To simulate the forward and backprojection operators, we use the Operator Discretization Library (ODL)~\cite{odl} with a 2D fan-beam geometry ($512$ detector pixels, source-to-axis distance of $600$ mm, axis-to-detector distance of $290$ mm). Sparse-view CT images are generated with $n_v \in \{32, 64, 128\}$ projection views, uniformly sampled from a full set of $512$ views covering $\left[0, 2\pi\right]$. To mimic real-world CT images, we introduce mixed noise to the sinograms, combining 5\% Gaussian noise and Poisson noise with an intensity of $1 \times 10^6$.

\seg{Training details.}
For each set of $n_v$ views, we train our model for $50$ epochs using $4$ Nvidia Tesla V100 (32GB RAM). We employ the AdamW optimizer~\cite{adamw} with a learning rate of \(1 \times 10^{-4}\), weight decay \(1 \times 10^{-2}\), and utilize the mean squared error loss with a batch size of $1$. Additionally, we incorporate a learning rate decay factor of $0.1$ after $40$ epochs. Unrolling iterations for \methodname are set to $T=14$. \regterm uses a patch size of $p=4$, $d=96$ embedding dimension, and $N=2$ mixer layers. The inverse Hessian size is $64^2 \times 64^2$ with $k=2$ downsampling blocks. $\mathcal{E}$ comprises cascading $3\text{x}3$ CNNs with max-pooling for downsampling, culminating in a $1\text{x}1$ CNN layer for a one-channel latent gradient. $\mathcal{D}$ utilizes 2x2 ConvTranspose operations. Both $\mathcal{E}$ and $\mathcal{D}$ layers incorporate instance normalization and PReLU activation. Following~\cite{regformer}, $\mA^{\dagger}$ is implemented using the FBP algorithm for the pseudo-inverse of $\mA$.

\seg{Evaluation metrics.}
Following established evaluation protocols \cite{learnedpd, dudotrans, regformer}, we employ the structural similarity index measure (SSIM) with parameters: level $5$, a Gaussian kernel of size $11$, and standard deviation $1.5$, as our primary performance metric. Furthermore, we supplement our assessment with the peak signal-to-noise ratio (PSNR).

\seg{State-of-the-art baselines.}\label{seg:sota-baselines} 
We compare \methodname to multiple state-of-the-art competitors: (1) \textit{post-processing} based denoising methods, i.e., FBPConvNet~\cite{fbpconvnet}, and DuDoTrans~\cite{dudotrans}; (2) \textit{first-order unrolling} reconstruction networks, i.e.,  Learned Primal-Dual~\cite{learnedpd}, LEARN~\cite{learn}, and RegFormer~\cite{regformer}. Note that we replace the pseudo-inverse operator used by LEARN with the FBP algorithm, as it has been demonstrated to be more effective according to~\cite{regformer}. To ensure a fair comparison, we utilize the code-base released by the authors when possible or meticulously implement the methods based on the details provided in their papers. All approaches undergo training and testing on the same datasets, as elaborated in \crefcustom{seg:imple_details}{implementation details}.

\subsection{Comparison with state-of-the-art methods}
\label{sec:exper_compar}

\input{tab/deeplesion_quantitative}

\seg{Quantitative comparison.}
We compared our model with \crefcustom{seg:sota-baselines}{state-of-the-art baselines} on two public datasets. For AAPM, models were trained and tested across three projection views ($n_v \in \{32, 64, 128\}$) and three noise levels, namely no noise $N_0=0$, low noise $N_1=10^6$, and high noise $N_2=5\times 10^5$ (see \cref{tab:aapm_quantitative}).
For DeepLesion, models were trained and tested on the same three projection views and a noise level of $N_1=10^6$ (see \cref{tab:deeplesion_quantitative}). 
Visual results are provided in \cref{fig:aapm_visual} (AAPM) and \cref{fig:deeplesion_visual} (DeepLesion).
Impressively, our method achieves state-of-the-art results on DeepLesion across all projection views. It outperforms the second-best baseline, RegFormer, with an average improvements of $+2.23$ dB in PSNR and $+1.02$\% in SSIM.
On AAPM without noise, we achieve state-of-the-art results across all projection views and improve the second best by an average $+1.65$ dB and $+0.58\%$.
In the presence of low noise, \methodname achieves state-of-the-art results performance in all cases except $n_v{=}128$ with $-0.11$ dB and shows an average improvements of $+0.33$ dB and $+0.35$\% over RegFormer.
With high noise, our method performs nearly on par in $n_v{=}32$ ($-0.02$ dB and $-0.29$\%), achieves state-of-the-art in $n_v{=}64$ ($+0.2$ dB and $+0.08$\%), and competes closely in $n_v{=}128$ ($-0.01$ dB and $+0.08$\%).
As noise increases, we attribute the decline in improvement to the compressed gradient information in the latent BFGS, influenced by sinogram changes, and the utilization of the FBP algorithm instead of the pseudo-inverse.

\seg{Performance comparison on OOD textures.}
\input{tab/ood_quantitative}
We evaluate frozen model performance on CT images featuring a randomly positioned white circle with noise-free sinograms, as illustrated in the third row of \cref{fig:aapm_visual}.
The rationale and details are provided in the supplementary material.
In \cref{tab:ood_quantitative}, \methodname attains state-of-the-art results across all $n_v$ views. First-order unrolling networks such as LEARN and RegFormer exhibit significant PSNR degradation of $-3.1$ dB and $-4.22$ dB, respectively, for $n_v{=}32$, while our method demonstrates a milder degradation of $-2.67$ dB.

\seg{Visual comparison.}
As it can be seen on \cref{fig:aapm_visual} and \cref{fig:deeplesion_visual}, FBPConvNet and DuDoTrans show significant blurring and pronounced artifacts when $n_v{=}32$. While Learned PD and LEARN show satisfactory performance, they struggle with intricate details, like in the liver and spine. In contrast, RegFormer produces high-quality images but faces challenges in generalizing to OOD data.
\methodname excels in producing high-quality images with fine details, even under challenging conditions such as $n_v{=}32$ views and OOD data.

\input{tab/costs_aapm}
\seg{Efficiency comparison.} \label{seg:efficiency}
The results in \cref{tab:costs_aapm} show that \methodname is more computationally efficient than RegFormer, with a $1.3\times$ reduction in memory usage.
Furthermore, our training time demonstrates a significant enhancement, realizing a speed improvement of $106$ seconds per epoch compared to first-order unrolling methods like LEARN and RegFormer. Additionally, our method requires only $14$ iterations, in contrast to the $30$ and $18$ iterations needed by LEARN and RegFormer, respectively.

\input{tab/ablation}

\subsection{Ablation Study}
\label{sec:exper_ablation}
In this section, we leverage the AAPM dataset with $n_v{=}32$ views by default, and no noise is introduced to the sinogram.

\seg{Inverse Hessian approximation size.}
The results in \cref{tab:hess_size} emphasize the significant impact of the inverse Hessian approximation size on our performance. When too small, a notable degradation is observed (e.g., $8^2 \times 8^2$), while larger sizes result in performance improvements as the approximation approaches the full inverse Hessian. Exceeding $64^2 \times 64^2$ was unfeasible in our experiments due to memory constraints.

\begin{figure}[htbp]
    \centering
    \captionsetup[subfigure]{labelformat=empty}
    \subfloat[]{
    \includegraphics[width=0.45\linewidth]{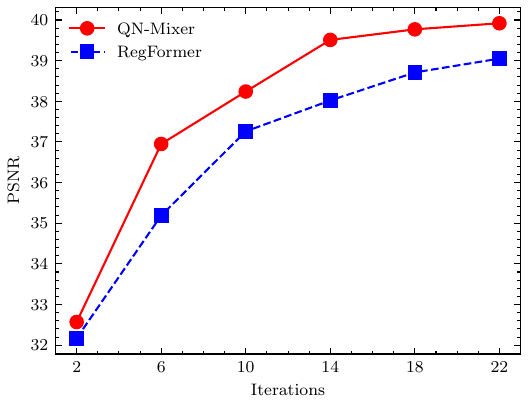}
    \label{subfig:abla_it_psnr}
    }
    \subfloat[]{
    \includegraphics[width=0.45\linewidth]{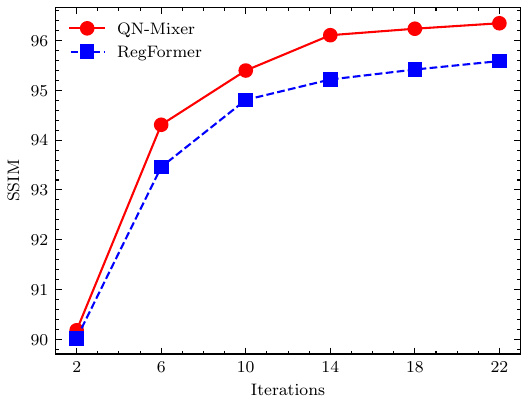}
    \label{subfig:abla_it_ssim}
    }
    \vspace{-0.5cm}
    \caption{\textbf{Ablation on the number of unrolling iterations}. 
    Left: PSNR (dB); Right: SSIM (\%)}
    \label{fig:ablation_qualitative}
    \vspace{-1.5em}
\end{figure}

\seg{Number of unrolling iterations.}
In \cref{fig:ablation_qualitative}, we visually depict the influence of the number of unrolling iterations on the performance of \methodname and RegFormer. Notably, the performance of both methods shows improvement with an increase in the number of iterations. When subjected to an equal number of iterations, our method consistently surpasses RegFormer in performance. Remarkably, we achieve comparable results to RegFormer even with only 10 iterations, demonstrating the efficiency of our approach.

\seg{Regularization term.}
In \cref{tab:reg_ablation}, we evaluate the impact of the regularization term in our framework.
Our \regterm is compared against various learned alternatives, including the Inception block \cite{inception} and MLP-Mixer block \cite{mlpmixer}. 
Additionally, employing the pseudo-inverse \textit{$\mA^{\dagger}$} instead of the FBP results in a less pronounced degradation ($-0.57$ dB and $-0.28$\%), enhancing the interpretability of \methodname. 
Finally, we test our \regterm in the first-order framework, highlighting the significance of the second-order latent BFGS approximation with a significant improvement ($+2.06$ dB and $+1.86\%$).

%% file: tab/aapm_quantitative.tex
\begin{table*}[htpb]
\centering
\setlength{\tabcolsep}{0.2em}
\resizebox{1.0\textwidth}{!}{
\begin{tabular}{lllllll@{\hskip 15pt}llllll@{\hskip 15pt}llllll}
\toprule
\multirow{3}{*}{\textbf{Method}} & \multicolumn{6}{c}{\textbf{No noise $(N_0=0)$}} & \multicolumn{6}{c}{\textbf{Low noise $(N_1=10^6)$}} & \multicolumn{6}{c}{\textbf{High noise $(N_2=5 \times 10^5)$}} \\
\cmidrule(lr){2-7}
\cmidrule(lr){8-13}
\cmidrule(lr){14-19}
{} & \multicolumn{2}{c}{\textbf{$n_v=32$}} & \multicolumn{2}{c}{\textbf{$n_v=64$}} & \multicolumn{2}{c}{\textbf{$n_v=128$}} & \multicolumn{2}{c}{\textbf{$n_v=32$}} & \multicolumn{2}{c}{\textbf{$n_v=64$}} & \multicolumn{2}{c}{\textbf{$n_v=128$}} & \multicolumn{2}{c}{\textbf{$n_v=32$}} & \multicolumn{2}{c}{\textbf{$n_v=64$}} & \multicolumn{2}{c}{\textbf{$n_v=128$}} \\
\cmidrule(lr){2-3}
\cmidrule(lr){4-5}
\cmidrule(lr){6-7}
\cmidrule(lr){8-9}
\cmidrule(lr){10-11}
\cmidrule(lr){12-13}
\cmidrule(lr){14-15}
\cmidrule(lr){16-17}
\cmidrule(lr){18-19}
 & {PSNR $\uparrow$} & {SSIM $\uparrow$} & {PSNR $\uparrow$} & {SSIM $\uparrow$} & {PSNR $\uparrow$} & {SSIM $\uparrow$} & {PSNR $\uparrow$} & {SSIM $\uparrow$} & {PSNR $\uparrow$} & {SSIM $\uparrow$} & {PSNR $\uparrow$} & {SSIM $\uparrow$} & {PSNR $\uparrow$} & {SSIM $\uparrow$} & {PSNR $\uparrow$} & {SSIM $\uparrow$} & {PSNR $\uparrow$} & {SSIM $\uparrow$} \\
\midrule
FBP & 22.65 & 40.49 & 27.29 & 57.94 & 33.04 & 79.50 & 22.09 & 32.73 & 26.51 & 49.56 & 31.69 & 71.09 & 19.05 & 15.56 & 22.71 & 25.74 & 26.52 & 40.87\\
FBPConvNet~\cite{fbpconvnet} & 30.32 & 85.11 & 35.42 & 90.15 & 39.71 & 94.64 & 30.20 & 84.46 & 35.09 & 89.72 & 39.06 & 94.08 & 29.91 & 82.52 & 34.13 & 87.85 & 36.89 & 91.28\\
DuDoTrans~\cite{dudotrans} & 30.48 & 84.70 & 35.37 & 91.87 & 40.62 & 96.41 & 30.34 & 83.72 & 35.36 & 91.42 & 39.75 & 95.49 & 30.09 & 81.83 & 34.09 & 88.67 & 37.08 & 93.44\\
Learned PD~\cite{learnedpd} & 35.88 & 92.09 & 41.03 & 96.28 & 43.33 & 97.31 & 35.78 & 92.21 & 39.03 & 94.79 & 41.65 & 96.44 & 33.80 & 89.23 & 37.34 & 93.23 & 39.17 & 94.69\\
LEARN~\cite{learn} & 37.58 & 94.65 & 42.26 & 97.25 & 43.11 & 97.57 & 36.95 & 93.63 & 39.91 & 95.82 & 42.17 & 97.11 & 34.38 & 90.51 & 37.15 & 93.53 & 39.38 & 95.18\\
RegFormer~\cite{regformer} & \underline{38.71} & \underline{95.42} & \underline{43.56} & \underline{97.76} & \underline{47.95} & \underline{98.98} & \underline{37.21} & \underline{94.73} & \underline{41.65} & \underline{96.92} & \textbf{44.38} & \underline{98.02} & \textbf{35.93} & \textbf{92.78} & \underline{38.53} & \underline{94.84} & \textbf{40.52} & \underline{96.19}\\
\cmidrule(l){1-19}
QN-Mixer~\textcolor{trolleygrey}{(ours)} & \textbf{39.51} & \textbf{96.11} & \textbf{45.57} & \textbf{98.48} & \textbf{50.09} & \textbf{99.32} & \textbf{37.50} & \textbf{94.92} & \textbf{42.46} & \textbf{97.70} & \underline{44.27} & \textbf{98.11} & \underline{35.91} & \underline{92.49} & \textbf{38.73} & \textbf{94.92} & \underline{40.51} & \textbf{96.27}\\
\bottomrule
\end{tabular}
}
\vspace{-.1cm}
\caption{\textbf{Quantitative evaluation on AAPM} of state-of-the-art methods (PSNR in dB and SSIM in \%). \textbf{Bold}: Best, \underline{under}: second best.}
\vspace{-0.5em}
\label{tab:aapm_quantitative}
\end{table*}

%% file: tab/deeplesion_quantitative.tex
\begin{table}
\centering
\setlength{\tabcolsep}{8pt}
\resizebox{.45\textwidth}{!}{
\begin{tabular}{lll@{\hskip 15pt}ll@{\hskip 15pt}ll}
\toprule
\multirow{2}{*}{\textbf{Method}} & \multicolumn{2}{c}{\textbf{$n_v  =  32$}} & \multicolumn{2}{c}{\textbf{$n_v  =  64$}} & \multicolumn{2}{c}{\textbf{$n_v  =  128$}} \\
\cmidrule(lr){2-3}
\cmidrule(lr){4-5}
\cmidrule(lr){6-7}
 & {PSNR $\uparrow$} & {SSIM $\uparrow$} & {PSNR $\uparrow$} & {SSIM $\uparrow$} & {PSNR $\uparrow$} & {SSIM $\uparrow$} \\
\midrule
FBP & 21.55 & 31.65 & 26.07 & 47.17 & 31.49 & 69.63\\
FBPConvNet~\cite{fbpconvnet} & 30.74 & 80.41 & 34.64 & 87.36 & 38.69 & 92.94\\
DuDoTrans~\cite{dudotrans} & 32.11 & 79.86 & 36.02 & 88.14 & 40.47 & 93.81\\
Learned PD~\cite{learnedpd} & 34.02 & 88.44 & 37.56 & 92.46 & 40.79 & 95.32\\
LEARN~\cite{learn} & 35.76 & 92.12 & 39.83 & 95.66 & 41.34 & 96.21\\
RegFormer~\cite{regformer} & \underline{37.38} & \underline{93.89} & \underline{41.70} & \underline{96.78} & \underline{46.10} & \underline{98.39}\\
\cmidrule(l){1-7}
QN-Mixer~\textcolor{trolleygrey}{(ours)} & \textbf{39.39} & \textbf{95.67} & \textbf{43.75} & \textbf{97.73} & \textbf{48.62} & \textbf{98.64}\\
\bottomrule
\end{tabular}
}
\caption{\textbf{Quantitative evaluation on DeepLesion} for state-of-the-art methods (PSNR in dB and SSIM in \%). With Poisson noise level of \(N_1=10^6\). \textbf{Bold}: Best, \underline{under}: second best.}
\vspace{-1em}
\label{tab:deeplesion_quantitative}
\end{table}

%% file: tab/ood_quantitative.tex
\begin{table}[t]
\centering
\setlength{\tabcolsep}{8pt}
\resizebox{.45\textwidth}{!}{
\begin{tabular}{lll@{\hskip 15pt}ll@{\hskip 15pt}ll}
\toprule
\multirow{2}{*}{\textbf{Method}} & \multicolumn{2}{c}{\textbf{$n_v=32$}} & \multicolumn{2}{c}{\textbf{$n_v=64$}} & \multicolumn{2}{c}{\textbf{$n_v=128$}} \\
\cmidrule(lr){2-3}
\cmidrule(lr){4-5}
\cmidrule(lr){6-7}
 & {PSNR $\uparrow$} & {SSIM $\uparrow$} & {PSNR $\uparrow$} & {SSIM $\uparrow$} & {PSNR $\uparrow$} & {SSIM $\uparrow$} \\
\midrule
FBP & 21.38 & 33.36 & 26.08 & 50.29 & 31.43 & 73.06\\
FBPConvNet~\cite{fbpconvnet} & 28.05 & 75.96 & 32.50 & 82.90 & 35.45 & 88.14\\
DuDoTrans~\cite{dudotrans} & 28.11 & 68.17 & 32.71 & 83.26 & 36.41 & 90.36\\
Learned PD~\cite{learnedpd} & 31.96 & 87.10 & 36.40 & 92.57 & 37.63 & 93.17\\
LEARN~\cite{learn} & 34.48 & \underline{90.15} & 36.89 & \underline{91.85} & \underline{38.32} & \underline{94.67}\\
RegFormer~\cite{regformer} & \underline{34.49} & 89.98 & \underline{36.95} & 91.48 & 38.02 & 92.44\\
\cmidrule(l){1-7}
QN-Mixer~\textcolor{trolleygrey}{(ours)} & \textbf{36.84} & \textbf{94.84} & \textbf{42.11} & \textbf{97.78} & \textbf{45.69} & \textbf{98.82}\\
\bottomrule
\end{tabular}
}
\caption{\textbf{Quantitative evaluation on out-of-distribution (OOD) AAPM test dataset} of state-of-the-art methods (PSNR in dB and SSIM in \%). \textbf{Bold}: Best, \underline{under}: second best.}
\vspace{-1em}
\label{tab:ood_quantitative}
\end{table}

%% file: tab/costs_aapm.tex
\begin{table}[htbp]
\centering
\setlength{\tabcolsep}{4pt}
\resizebox{.45\textwidth}{!}{
\begin{tabular}{lccccc}
\toprule
Method & \#Iters & Epoch time (s) & Time (ms) & \#Params (M) & Memory (GB) \\
\toprule
\subtabtitle{Post-processing based denoising}{6} \\
FBPConvNet~\cite{fbpconvnet} & - & 68 & 12.4 & 31.1 & 1.30\\
DuDoTrans~\cite{dudotrans} & - & 92 & 60.1 & 15.0 & 1.38\\
\midrule
\subtabtitle{First-order unrolling reconstruction networks}{6} \\
Learned PD~\cite{learnedpd} & 10 & 82 & 47.2 & 0.25 & 0.81\\
LEARN~\cite{learn} & 30 & 780 & 679.8 & 4.50 & 1.85\\
RegFormer~\cite{regformer} & 18 & 700 & 598.9 & 5.00 & 10.19\\
\midrule
\subtabtitle{Second-order unrolling Quasi-Newton}{6} \\
QN-Mixer~\textcolor{trolleygrey}{(ours)} & 14 & 594 & 610.2 & 8.50 & 7.83\\
\bottomrule
\end{tabular}
}
\vspace{-0.2cm}
\caption{\textbf{Comparison of computational efficiency}. Training epoch time is reported in seconds, \#Params in M and memory costs for state-of-the-art methods on AAPM with $n_v=32$ views.}
\label{tab:costs_aapm}
\end{table}

%% file: tab/ablation.tex
\begin{table}[htbp]
\begin{minipage}{0.42\linewidth}    
    \centering
    \resizebox{\textwidth}{!}{
    \begin{tabular}{lc@{\hskip 15pt}c}
        \toprule
        Hessian size & {PSNR $\uparrow$} & {SSIM $\uparrow$}\\
        \midrule
        $8^2 \times 8^2$ & 35.69 & 93.71\\
        $16^2 \times 16^2$ & 38.11 & 95.31\\
        $32^2 \times 32^2$ & 39.37 & 96.01\\
        \hdashline[2pt/4pt]
        \textbf{$64^2 \times 64^2$} & \textbf{39.51} & \textbf{96.11}\\
        \bottomrule
    \end{tabular}
    }
    \captionof{table}{\textbf{Ablation on the inverse Hessian approximation size}.}
    \label{tab:hess_size}
\end{minipage}\hfill
\begin{minipage}{0.53\linewidth}
    \resizebox{\linewidth}{!}{
    \begin{tabular}{lc@{\hskip 15pt}c}
        \toprule
        Method & {PSNR $\uparrow$} & {SSIM $\uparrow$}\\
        \midrule
        \subtabtitle{QN with different learned regularization}{3} \\
        Inception & 31.65 & 85.28\\
        U-Net & 34.29 & 92.92\\
        MLP-Mixer & 36.89 & 93.87\\
        Incept-Mixer & 39.51 & 96.11 \\
        \midrule
        \subtabtitle{Pseudo-inverse $\mA^{\dagger}$ vs Filtered Back Projection (FBP)}{3} \\
        QN-Mixer+$\mA^{\dagger}$ & 38.94 & 95.83\\
        QN-Mixer+FBP & 39.51 & 96.11\\
        \midrule
        \subtabtitle{First vs second order Quasi-Newton (QN)}{3} \\
        Incept-Mixer+first-order & 37.45 & 94.25\\
        Incept-Mixer+QN & 39.51 & 96.11\\
        \bottomrule
        \end{tabular}
    }
    \captionof{table}{
        \textbf{\methodname ablation}.
    }
    \label{tab:reg_ablation}
\end{minipage}
\vspace{-1.6em}
\end{table}

%% file: sec/5_conclusion.tex
\section{Conclusion}
\label{sec:conclu}
In this paper, we investigate the application of deep second-order unrolling networks for tackling imaging inverse problems. To this end, we introduce \methodname, a quasi-Newton inspired algorithm where a latent BFGS method approximates the inverse Hessian, and our \regterm serves as the non-local learnable regularization term. 
Extensive experiments confirm the successful sparse-view CT reconstruction by our model, showcasing superior performance with fewer iterations than state of-the-art methods. 
In summary, this research offers a fresh perspective that can be applied to any iterative reconstruction algorithm. 
A limitation of our work is the memory requirements associated with quasi-Newton algorithm. We introduced a memory efficient alternative by projecting the gradient to a lower dimension, successfully addressing the CT reconstruction problem. However, its applicability to other inverse problems may be limited. 
In future work, we aim to extend our approach to handle larger Hessian sizes, broadening its application to a range of problems.

%% file: sec/X_suppl.tex
\clearpage
\setcounter{page}{1}
\setcounter{section}{0}
\maketitlesupplementary
\appendix

\startcontents{
    \hypersetup{linkcolor=black}
    \printcontents{}{1}{}
}

\noindent\rule{\linewidth}{0.4pt}

\section{Inverse Hessian approximation}
\label{sec:apx_hessian}
\input{apx/fig/hessian_fig}
\subsection{Approximation via optimization}
\label{sec:apx_hess_approx}
The fundamental idea is to iteratively build a recursive approximation by utilizing curvature information along the trajectory. 
It is crucial to emphasize that a quadratic approximation offers a direction that can be leveraged within the iterative update scheme. This direction is defined by the equation:

\begin{equation}
\vx_{t+1}=\vx_t + \alpha_t \vd_t.
\label{eq:update}
\end{equation}

In order to determine the direction $\vd_t$, we can employ a quadratic approximation of the objective function. This approximation can be expressed as:
{\small
\begin{equation}
J(\vx_t + \vd) \approx m_t(\vd)=J(\vx_t) + \nabla J(\vx_t)^{\mathsf{T}} \vd + \frac{1}{2} \vd^{\mathsf{T}} \mB_t \vd,
\label{eq:quaddirection}
\end{equation}
}

\noindent where $J(\vx_t)$ represents the objective function evaluated at the current point $\vx_t$, $\nabla J(\vx_t)$ denotes the gradient of the objective function at $\vx_t$, $\mB_t \in \R^{n \times n}$ corresponds to the approximation of the Hessian matrix. By minimizing the right-hand side of the quadratic approximation in \cref{eq:quaddirection}, we can determine the optimal direction $\vd_t$. Taking the derivative of $m_t(\vd)$ with respect to $\vd$ and setting it to zero, we obtain:
{\small
\begin{equation*}
\frac{\nabla m_t(\vd)}{\nabla \vd}=\vd_t \mB_t + \nabla J(\vx_t) \xrightarrow{\nabla m_t(\vd)=0} \vd_t=-\mB_t^{-1} \nabla J(\vx_t),
\end{equation*}
}

\noindent by substituting this result in \cref{eq:update} we obtain:
\begin{equation*}
    \vx_{t+1}=\vx_t - \alpha_t \mB_t^{-1} \nabla J(\vx_t).
\end{equation*}

The objective is to ensure that the curvature along the trajectory is consistent. In other words, at the last two iterations, $m_{t+1}$ should match the gradient $\nabla J(\vx_t)$ in the following way:
\begin{equation*}
\nabla m_{t+1} \bigg|_{\vd=0}=\nabla J(\vx_{t+1}),\
\nabla m_{t+1} \bigg|_{\vd=-\alpha_t \vd_t}=\nabla J(\vx_{t}).
\end{equation*}

This condition ensures that the quadratic approximation captures the correct curvature information along the trajectory, allowing for accurate optimization and convergence of the algorithm. By evaluating $\nabla m_{t+1}(\cdot)$ at the point $-\alpha_t \vd_t$ we obtain:
\begin{equation*}
    \alpha_t \mB_{t+1} \vd_t=\nabla J(\vx_{t+1}) - \nabla J(\vx_{t}).
\end{equation*}

From \cref{eq:update} we get the secant equation:
{\small
\begin{equation*}
    \mB_{t+1} {\color{softgreen} \underbrace{\text{$(\vx_{t+1} - \vx_{t})$}}_\text{$\vs_t$}}={\color{softblue}\underbrace{\text{$\nabla J(\vx_{t+1}) - \nabla J(\vx_{t})$}}_\text{$\vz_t$}} \rightarrow \mB_{t+1}{\color{softgreen} \text{$\vs_t$}}={\color{softblue} \text{$\vz_t$}}.
\end{equation*}
}

To avoid explicitly computing the inverse matrix $\mB_t^{-1}$, we can introduce an approximation $\mH_t=\mB_t^{-1}$ and optimize it as follows:
\begin{equation}
    \begin{aligned}
    \mH_{t+1}=\argmin{\mH} \quad & \norm{\mH - \mH_t}^2_\mW & \rhd \mH_{t+1}\ \text{close to}\ \mH_t \\
    \text{s.t.:} \quad & \mH=\mH^{\mathsf{T}} & \rhd \ \text{symmetry} \\
    & \mH \vz_t=\vs_t & \rhd \text{secant equation}
    \end{aligned}
    \label{eq:h_optim}
\end{equation}

\noindent Here $\norm{\cdot}_\mW^2$ denotes the weighted Frobenius norm. This optimization problem aims to find an updated approximation $\mH_{t+1}$ that is close to $\mH_t$, while satisfying the constraints that $\mH_{t+1}$ is symmetric and satisfies the secant equation $\mH_{t+1}\vz_t=\vs_t$. BFGS~\cite{bfgs, qnbfgs, qn} uses $\mW=\int_0^1 \nabla^2 J(\vx_{t} + t \alpha_t \vd_t) dt$, to solve this optimization problem and obtain the iterative update of $\mH$:
\begin{equation}
\mH_{t+1}=(\mI - \rho_t \vs_t \vz_t^\mathsf{T}) \mH_t (\mI - \rho_t \vz_t \vs_t^\mathsf{T}) + \rho_t \vs_t \vs_t^\mathsf{T},    
\end{equation}

\noindent where $\rho_t=\frac{1}{\vz_t^{\mathsf{T}}\vs_t}$, $\vs_t=\vx_{t+1} - \vx_t$, and $\vz_t=\nabla J(\vx_{t+1}) - \nabla J(\vx_t)$. This update equation allows us to iteratively refine the approximation $\mH_t$ based on the current gradient information and the changes in the solution. 
\input{apx/fig/hess_req}

\begin{figure*}[t] 
    \centering 
    \includegraphics[width=1.0\textwidth]{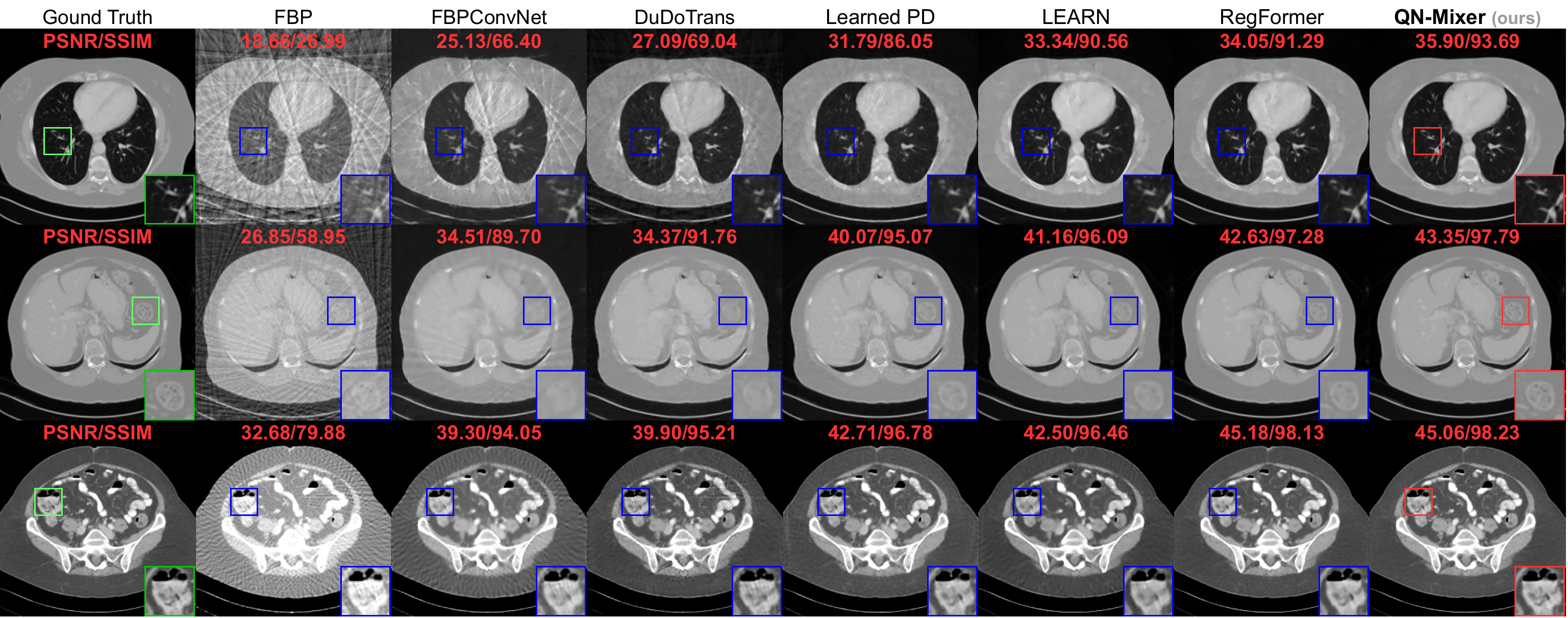}
    \vspace{-0.5cm}
    \caption{\textbf{Visual comparison on AAPM}. From top to bottom: the results under the following conditions: first $(n_v=32, N_1)$, second $(n_v=64, N_1)$, third $(n_v=128, N_1)$. The display window is set to $\left[-1000, 800\right]$ HU for the first two rows and to $\left[-200, 300\right]$ HU for the last row.}    
    \label{fig:apx_aapm_visual} 
    \vspace{-0.1em}
\end{figure*}

\subsection{Validation of adherence to BFGS}
\label{sec:apx_validation_bfgs}
We validate the adherence of our method to the BFGS requirements. To achieve this, we present the constraint values of the optimization algorithm given in \cref{eq:h_optim} using a test set image from AAPM, as illustrated in \cref{fig:hess_req}. The symmetry index is defined as follows:
{\small
\begin{equation}
    SI=\frac{1}{{n \cdot (n-1)}} \sum_{i=1}^{n} \sum_{j=1, j \neq i}^{n} \lvert \mA_{ij} - \mA_{ji} \rvert .
    \label{eq:symidx}
\end{equation}
}
Our results demonstrate the effectiveness of our approach in satisfying the essential conditions required by the BFGS algorithm. Notably, the symmetry index is consistently close to zero, indicating the symmetry of the matrix $\mH_t$ at each iteration, which is the first constraint of the BFGS method. Furthermore, with regard to the objective function value, it is evident that it is close to zero, except for the initial approximation. This deviation can be attributed to the use of the identity matrix as the starting point.

\paragraph{Inverse Hessian matrix approximation visualizations.}
\Cref{fig:hess_visual} depicts $\mH_t$ at different iterations. These visualizations confirm the required symmetry of the matrix in each iteration. Additionally, the matrix $\mH_t$ is close to the identity matrix at the second iteration, becoming more structured in the third iteration. This behavior aligns with expectations, as the matrix $\mH_t$ is initialized as the identity matrix and updated based on gradient information and solution changes. 

\paragraph{Visualization of reshaped rows.}
To further understand the inverse Hessian matrix approximation structure, we depict the reshaped ($64 \times 64$) first $40$ rows of $\mH_t$ at iterations $9$ and $14$ in \cref{fig:hess_visual_rows}. These rows store gradient attention information used for updating the solution, consistent with the matrix $\mH_t$ being updated based on gradient information and solution changes. In future work, we plan to explore the impact of $\mH_t$ on the optimization process and its influence on reconstruction performance.

\section{More Ablation Study and visualization}
\label{sec:apx_ablation}

\begin{figure*}[t] 
    \centering 
    \includegraphics[width=1.0\textwidth]{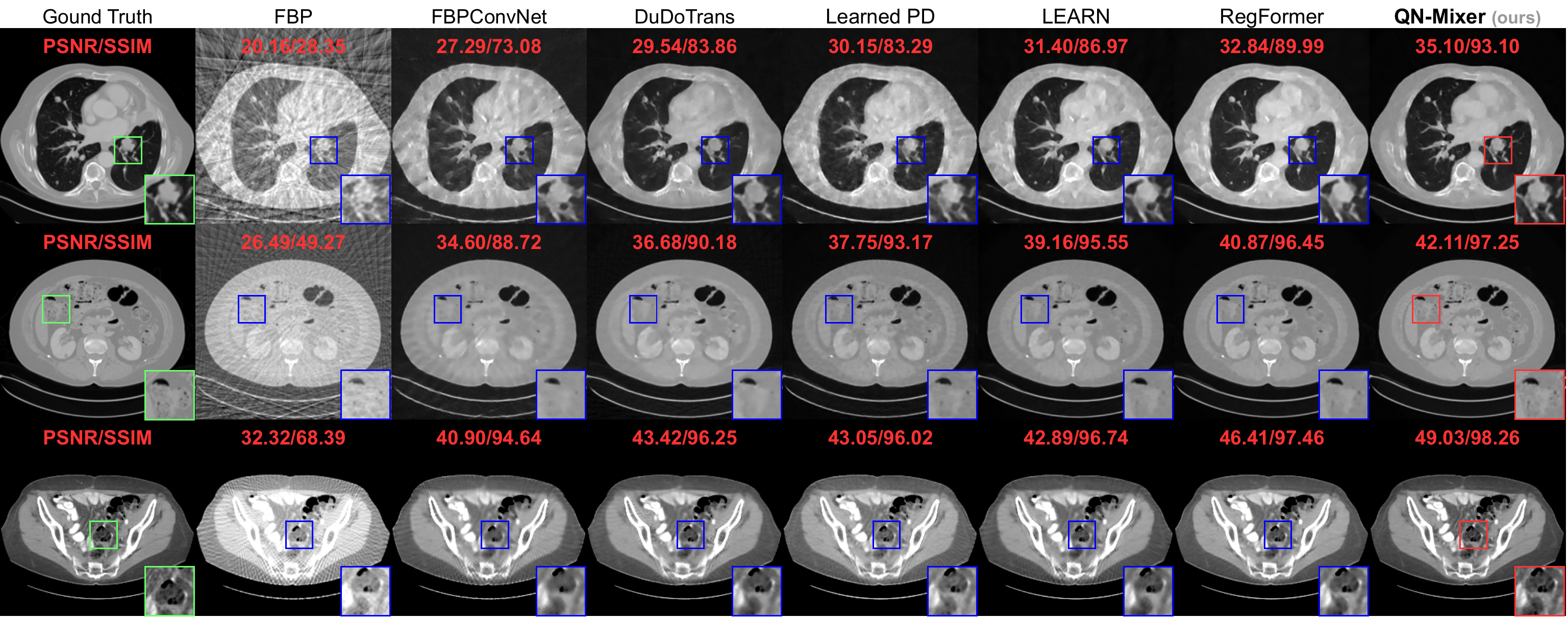}
    \vspace{-0.5cm}
    \caption{\textbf{Visual comparison on DeepLesion}. From top to bottom: the results under the following conditions: first $(n_v=32, N_1)$, second $(n_v=64, N_1)$, third $(n_v=128, N_1)$. The display window is set to $\left[-1000, 800\right]$ HU for the first two rows and to $\left[-200, 300\right]$ HU for the last row.}
    \label{fig:apx_deeplesion_visual}
\end{figure*}

\subsection{Ablation on \regterm}
\label{sec:apx_ablation_regterm}
We further investigate the impact of the hyperparameters of \regterm on the reconstruction performance. We vary the patch size $p$ and the number of stacked Mixer layers $N$ and report the results in \cref{tab:reg_patch}, and \cref{tab:reg_depth} respectively.

\paragraph{Impact of the path size.}
We observe that increasing the patch size $p$ from $2$ to $4$ improves the performance ($+1.28$ dB and $+1.04$\%) while further increasing the patch size from $4$ to $8$ decreases the performance ($-1.32$ dB and $-0.79$\%).

We attribute this observed pattern to the trade-off between local and global features in the reconstruction process. When the patch size is small, such as $p=2$, the model focuses on capturing fine-grained local details, which can enhance reconstruction accuracy. As the patch size increases to $p=4$, the network gains a broader perspective by considering larger regions, leading to an improvement in performance. However, when the patch size becomes too large, for example, $p=8$, the model might start incorporating more global context at the expense of losing finer details. This can result in a decrease in performance as the model becomes less sensitive to localized patterns.
\input{apx/tab/abx_ablation}

\paragraph{Impact of the number of stacked Mixer layer.}
We observe that increasing the number of stack $N$ from $1$ to $2$ improves the performance ($+1.86$ dB and $+1.31$\%), while further increasing the patch size from $2$ to $3$ decreases the performance ($-1.03$ dB and $-0.60$\%) and from $3$ to $4$ decreases the performance ($-0.30$ dB and $-0.11$\%).

Similarly, when varying the number of stacked Mixer layers $N$, we observe a trend where an increase in $N$ initially contributes to improved performance, as the model can capture more complex features and relationships.
However, as $N$ continues to grow, the network may encounter diminishing returns, and the benefits of additional layers diminish, potentially leading to overfitting or increased computational overhead.

\paragraph{Robustness to hyperparameters.}
Hence, there exists an optimal trade-off between the patch size $p$ and the number of stacked Mixer layers $N$, but the model performs similarly for a wide range of values. In our experiments, we use $p=4$ and $N=2$ for all the datasets, which highlights the robustness of our method to these hyperparameters.

\subsection{More visualization results}
\cref{fig:apx_aapm_visual} displays supplementary visualizations of our approach on AAPM. Our method consistently produces high-quality reconstructions across all views. Notably, among state-of-the-art techniques, \methodname excels in reconstructing fine-grained details. For instance, it accurately captures small vessels in the first row, delicate soft tissue structures in the second row, and sharp boundaries in the third row.

In \cref{fig:apx_deeplesion_visual}, we showcase additional visualizations of our method applied to DeepLesion. \methodname demonstrates superior performance across all views, yielding high-quality reconstructions. This is particularly evident in the challenging scenario of $32$ views, where our method outperforms others in capturing fine-grained details, such as small vessels and lesions. Importantly, these results are achieved with fewer iterations compared to alternative unrolling networks like RegFormer.

\subsection{Iterative results visualization}
\input{apx/fig/it_fig}
In order to demonstrate the effectiveness of \methodname, we present a series of intermediate reconstruction results in \cref{fig:it_fig}. These results illustrate the progression of the reconstruction process at different iterations of our method. By examining the reconstructed outputs at each iteration, our goal is to offer insights into the evolution of image quality. Notably, we observe that the improvement in quality, as quantified by the PSNR and SSIM values of each iteration, does not consistently increase with each iteration (see Iteration $10$ in \cref{fig:it_fig}).
We suspect that the observed unexpected behavior may arise from the variation of the objective function (i.e. \cref{eq:objective}) around the point $t$ in the unrolled network, which is dependent on a learnable gradient regularization term \regterm.

\subsection{Reconstruction error visualization}
\input{apx/fig/reconstruction}
We present the reconstruction error of \methodname in comparison to LEARN and RegFormer in \cref{apx:fig_reconstruction}. The images are organized from left to right based on the SSIM value. As illustrated, our method consistently produces high-quality reconstructions. In the most challenging scenario ($n_v=32$) with no added noise, and for the least favorable image, our method achieves a reconstruction with an SSIM of $93.53\%$, maintaining notably satisfactory performance compared to RegFormer with an SSIM of $91.30\%$. For the best reconstruction across all methods, our method achieves an SSIM of $97.72\%$, while RegFormer achieves an SSIM of $97.48\%$. These results demonstrate the robustness of our method when dealing with challenging scenarios.

\input{apx/tab/ood_patient}
\section{More experiments}

\subsection{Out-of-Distribution}
\input{apx/fig/ood_fig}

\seg{OOD circle performance across entire image}
In the main text, we assess the robustness of methods to a simple out-of-distribution scenario, where an unseen during training, white circle is inserted, and computing SSIM and PSNR metrics for the \emph{entire image}. We achieved the best performance across all views in this setting (Tab. 3). We note, however that this setting mixes the inherent performance on clean data and its ability to handle unseen patterns.

\seg{OOD circle performance in anomalous region}
To further isolate the capacity of models to reconstruct unseen OOD patterns (i.e., white circle), we extend our evaluation. Instead of evaluating the whole image, we compute the reconstruction performance of a crop \emph{region containing the white circle}, thus isolating the reconstruction performance exclusively to the circle region.

For that, we randomly selected 5 samples from the AAPM test set depicted. The evaluated data is depicted in \cref{apx:ood_fig}, and the overall performance across the complete set of 214 patient images is summarized in \cref{apx:ood_patient} for 3 views.

Our method significantly outperforms the second-best across all views in both SSIM and PSNR. For the most challenging case of $32$ views, we surpass the second best by $+6.47\%$ and $+1.49$ dB. With $64$ views, our performance exceeds the second best by $+10.12\%$ and $+4.21$ dB. In the easiest case of $128$ views, we outperform the second best by $+5.29\%$ and $+5.76$ dB. As anticipated, all methods exhibit degraded performance when focusing on the circle region, and the gap between our method and the second-best widens compared to the complete image. Moreover, the numerical results in \cref{apx:ood_patient} align with the visualizations in \cref{apx:ood_fig}.

\input{apx/tab/ood_hard}
\seg{Anatomy and Geometry OOD.}
We explore more complex out-of-distribution (OOD) scenarios, particularly focusing on changes in anatomy and geometry.
For the geometry aspect, we train our methods using AAPM data and evaluate them on DeepLesion, maintaining a fixed geometry parameter of $n_v{=}32$.
Concerning anatomy, we test under the most challenging scenario, which involves training on AAPM data with $n_v{=}128$ and subsequently testing on AAPM data with two lower geometries: $n_v{=}32$ and $n_v{=}64$.
The outcomes of this experiment are detailed in \cref{apx:ood_hard}.
\begin{itemize}
    \item \textbf{anatomy}: we outperform RegFormer by $+0.31$ dB and $+0.65\%$. 
    \item \textbf{geometry $n_v{=}32$}: we outperform Regformer by $+0.60$ dB and $+7.50\%$ in most challenging setting.
    \item \textbf{geometry $n_v{=}64$}: we lag behind Regformer by $-0.62$ dB and $-0.83\%$ in the easiest setting.
\end{itemize}

\subsection{Results when A is the $90^{\circ}$ limited view CT}
\begin{figure}[htbp]
    \centering
    \includegraphics[width=\linewidth]{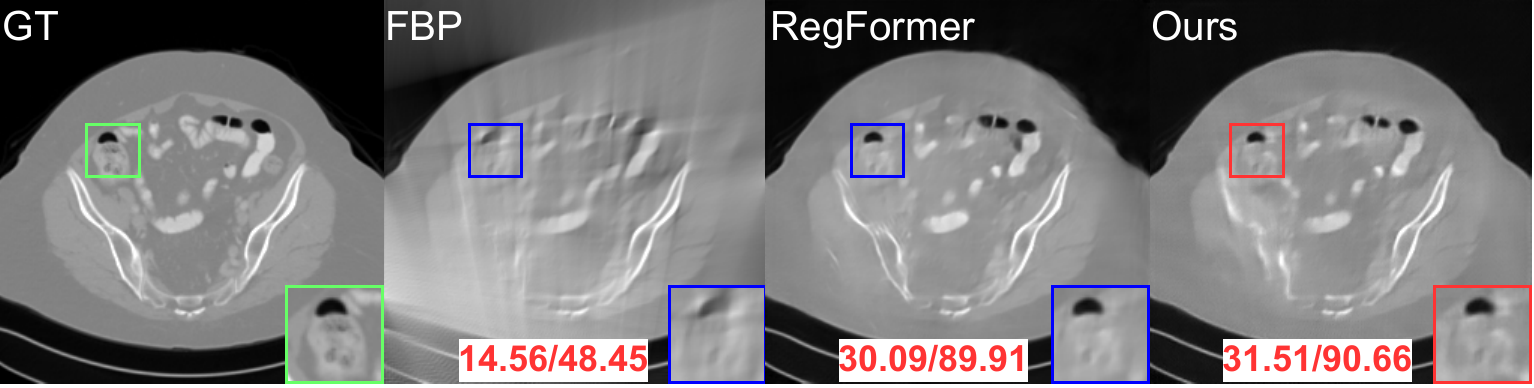}
    \caption{\textbf{Visual comparaison on AAPM with Limited-angle CT}. The system matrix $A$ is a $90^{\circ}$ limited view. The display window is set to $\left[-1000, 800\right]$ HU.}
    \label{fig:limited_angle}
\end{figure}

To further evaluate the robustness of our method, we conducted experiments on a more challenging CT inverse problem, specifically involving a $90^{\circ}$ limited-angle CT setup. In this scenario, the system matrix $A$ comprises a restricted number of views, resulting in significant artifacts and distortions in the reconstructed images. We compared the performance of our proposed \methodname, with that of RegFormer on this demanding task. Our method demonstrated superior performance, achieving promising results with a PSNR of $30.17$ dB and a SSIM of $90.01\%$. This outperformed RegFormer by $+1.03$ dB in PSNR and $+0.45\%$ in SSIM. Visual results illustrating the effectiveness of our approach are provided in \cref{fig:limited_angle}.

\subsection{MBIR evaluation}
To demonstrate the superiority of our method, we conducted an additional comparison with the state-of-the-art model-based iterative reconstruction (MBIR) technique, which is widely employed in clinical settings. Our evaluation involved testing our approach against MBIR using datasets with varying numbers of views, specifically $n_v{=}32, 64$, and $128$, with low noise added to sinograms. The corresponding PSNR values obtained with MBIR were $22.94$, $28.80$, and $34.33$ dB, with SSIM values of $67.70\%$, $72.60\%$, and $80.68\%$, respectively. 

Our method, \methodname, consistently outperformed MBIR across all views, showcasing an average improvement of $+12.72$ dB in PSNR and $+23.25\%$ in SSIM. These results underscore the robustness and effectiveness of our approach compared to MBIR, a benchmark technique widely utilized in clinical practice.

\subsection{Noise Power Spectrum analysis}
\input{apx/fig/nps}
\input{apx/fig/nps_roi}
We conducted a comprehensive examination of the noise characteristics in our reconstructed images through noise power spectrum (NPS) analysis. NPS serves as a metric, quantifying the magnitude and spatial correlation of noise properties, or textures, within an image. It is derived from the Fourier transform of the spatial autocorrelation function of a zero-mean noise image.

NPS analysis was performed on a configuration of Regions of Interest (ROIs) as depicted in \cref{fig:nps_roi}. This process was applied to all $214$ images from the AAPM test set and for three different views ($32$, $64$, and $128$). The average 1D curves were generated by radially averaging the 2D NPS maps, and the results are presented in \cref{apx:nps}.

The area under the NPS curve is equal to the square of the noise magnitude. Importantly, the ordering of methods based on noise magnitude corresponds to the ranking observed in our quantitative experiments for PSNR and SSIM in the main text. For example, FBP, which exhibits the lowest noise magnitude, also performs the poorest in terms of PSNR and SSIM. Conversely, our method, with the highest noise magnitude, stands out as the top performer in both PSNR and SSIM metrics. Furthermore, the mean and peak frequencies serve as key indicators of noise texture or ``noise grain size'', where higher frequencies denote finer texture. Remarkably, our method showcases superior mean and peak frequencies compared to other methods, suggesting a finer noise texture or smaller grain size.

This alignment with good clinical practice standards reinforces the robust performance of our method in capturing and preserving image details, as supported by both quantitative metrics and noise analysis.

\section{Reproducibility}
\input{apx/alg/qn_mixer}
All our experiments are fully reproducible. While the complete algorithm is already provided in the main paper (see \cref{algo:qnmixer}), we additionally present a PyTorch pseudo-code for enhanced reproducibility in \cref{apx:repro_qn_mixer}. We furnish comprehensive references to all external libraries used in \cref{apx:repro_lib}. Detailed information regarding the initialization of our model can be found in \cref{apx:repro_qn_init}. The precise parameters of our regularizer, \regterm architecture, are available in \cref{apx:repro_incept_mixer}. We outline the exact data splits utilized across the paper for the AAPM dataset in \cref{apx:repro_aapm}. Lastly, to facilitate the reproduction of our out-of-distribution (OOD) protocol, we provide the pseudo-code in \cref{apx:repro_ood}.

\subsection{\methodname pseudo-code} \label{apx:repro_qn_mixer}
Our \methodname algorithm is introduced in \cref{algo:qnmixer}, and for improved reproducibility, we present a PyTorch pseudo-code in \cref{apx:qn_mixer}. The fundamental concept underlying unrolling networks lies in having a modular gradient function, denoted as $\nabla J(\vx_t)$, which can be easily adapted to incorporate various regularization terms. Subsequently, the core element is the unrolling iteration block responsible for updating both the solution $\vx_t$ and the inverse Hessian approximation $\mH_t$. The update of the inverse Hessian approximation is executed through the latent BFGS algorithm. Notably, each iteration call takes the physics operator as input, tasked with computing the forward and pseudo-inverse operators for the CT reconstruction problem, along with the gradient encoder and direction decoder, which are shared across all iterations. For a more in-depth understanding, refer to \cref{apx:qn_mixer}. Note the employment of \texttt{torch.no\_grad()} to inhibit the computation of gradients for the inverse Hessian approximation. Since there is no necessity to compute gradients for this variable, given that it is updated through the latent BFGS algorithm.

Within these two modules, second-order quasi-Newton methods can be seamlessly incorporated by simply modifying the latent BFGS algorithm or the regularization term, offering flexibility to the user.

\subsection{External libraries used} \label{apx:repro_lib}
We utilized the following external libraries to implement our framework and conduct our experiments:
\begin{itemize}
    \item Operator Discretization Library (ODL): \\ {\footnotesize \url{https://github.com/odlgroup/odl}}
    \item High-Performance GPU Tomography Toolbox (ASTRA): \\ {\footnotesize \url{https://www.astra-toolbox.com/}}
    \item Medical Imaging Python Library (Pydicom): \\ {\footnotesize \url{https://pydicom.github.io/}}
\end{itemize}

\subsection{\methodname's parameters initialization} \label{apx:repro_qn_init}
To enhance reproducibility, we provide the parameters initialization of \methodname. \emph{First}, for the gradient function, we initialize the CNNs of \regterm using the Xavier uniform initialization. The multi-layer perceptron of the MLP-Mixer is initialized with values drawn from a truncated normal distribution with a standard deviation of $0.02$. The $\lambda_t$ values are initialized to zero, and the inverse Hessian approximation $\mH_0$ is initialized with the identity matrix $\mI$. \emph{Second}, for the latent BFGS, both the encoder and decoder CNNs are initialized with the Xavier uniform initialization.

\subsection{\regterm's architecture} \label{apx:repro_incept_mixer}
For enhanced reproducibility, we present the architecture of \regterm in \cref{tab:abx_arch}. The \regterm architecture consists of a sequence of Inception blocks, followed by Mixer blocks. Each Mixer block comprises a channel-mixing MLP and a spatial-mixing MLP. The MLPs are constructed with a fully-connected layer, a GELU activation function, and another fully-connected layer. Ultimately, the regularization value is projected to the same dimension as the input image through a patch expansion layer, which is composed of a fully-connected layer and a CNN layer.
\input{apx/tab/aapm_splits}
\input{apx/tab/abx_arch}

\subsection{AAPM dataset splits} \label{apx:repro_aapm}
In our experiments, we use the AAPM 2016 Clinic Low Dose CT Grand Challenge public dataset~\cite{aapm}, which holds substantial recognition as it was formally established and authorized by the esteemed Mayo Clinic. To ensure the integrity of our evaluation process, we followed the precedent set by~\cite{learn, regformer} and created the training set using data from eight patients, while reserving a separate patient for the testing and validation sets. This approach guarantees that no identity information is leaked during test time. Our specification is presented in \cref{tab:aapm_splits}.

\subsection{Robustness eval. protocol for OOD scenarios} \label{apx:repro_ood}
\input{apx/alg/circle}

In medical imaging, it's crucial to develop methods that generalize to scans with lesions or anomalies, and assessing the model's capability to reconstruct abnormal data holds significant relevance, as test patient data may deviate from the training data in clinical applications.
To this end, we design a simple protocol specifically crafted for evaluating the effectiveness of methods when handling abnormal data. In this case, a white circle mimicking an out-of-distribution texture, which was never seen during training, is forged into CT images with noise-free sinograms.
The pseudo-code to realize this is provided in \cref{apx:circle}.
Then, performance can be evaluated on the entire image as shown in \cref{tab:ood_quantitative}, or on a cropped region within the circle as detailed in \cref{apx:ood_patient}.

We strongly advocate for future research endeavors to embrace and employ this protocol as a standard for evaluating the robustness of reconstruction methods. 

\section{Limitations}
\label{sec:limitations}
Our approach inherits similar limitations from prior methods~\cite{learn, regformer}.
First, our method entails a prolonged optimization time, stemming from the utilization of unrolling reconstruction networks~\cite{learn, regformer}, in contrast to post-processing-based denoising methods~\cite{fbpconvnet, dudotrans}.
While our method represents the fastest unrolling network, there is still a need to address the existing gap.
Integrating Limited-memory BFGS into our \methodname framework is an interesting research direction for accelerating training. 
Second, while we have assessed our method using the well known AAPM low-dose and DeepLesion datasets and compared it with several state-of-the-art methods, the evaluation is conducted on images representing specific anatomical regions (thoracic and abdominal images). The generalizability of our method to a broader range of datasets, which may exhibit diverse characteristics or variations, remains unclear.
Third, the acquisition of paired data has always been an important concern in clinic.
Combining our approach with unsupervised training framework to overcome this limitation can be an exciting research direction. 
Finally, the incorporation of actual patient data into our training datasets raises valid privacy concerns. Although the datasets we utilized underwent thorough anonymization and are publicly accessible, exploring a solution that can effectively operate with synthetic data emerges as an intriguing avenue to address this challenge.

\section{Notations}
\label{sec:notations}
We offer a reference lookup table, available in \Cref{apx:notations}, containing notations and their corresponding shapes as discussed in this paper.

\input{apx/tab/notations}

%% file: apx/fig/hessian_fig.tex
\begin{figure*}[t] 
    \centering 
    \includegraphics[width=1.0\textwidth]{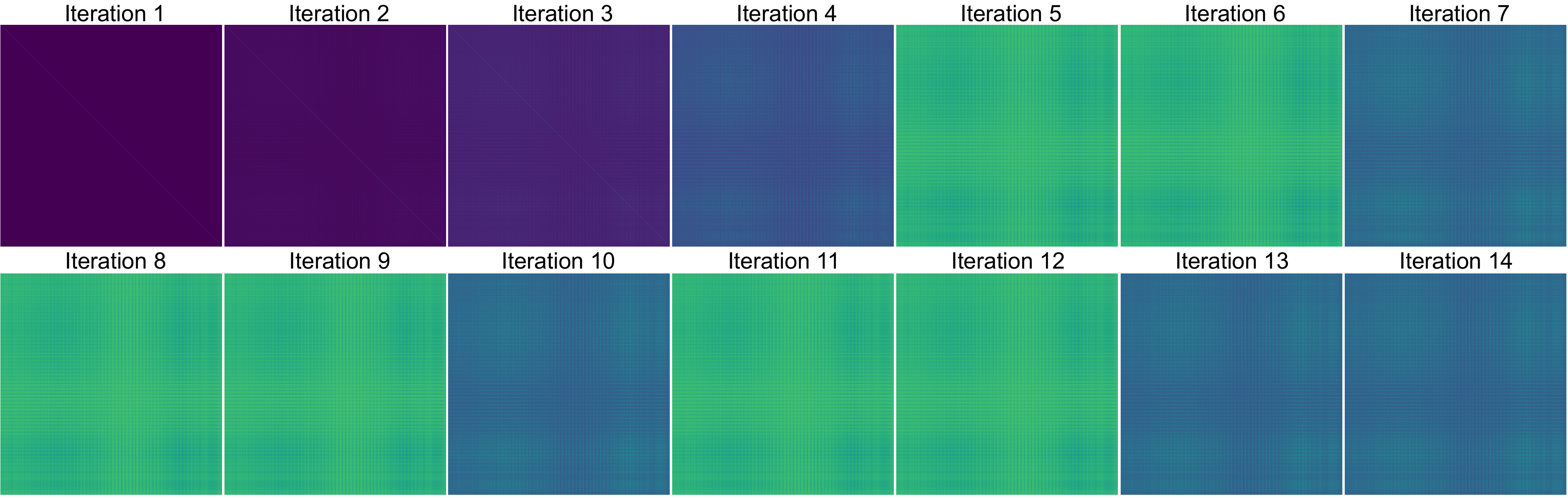}
    \vspace{-0.5cm}
    \caption{\textbf{Visualization of the inverse Hessian approximation across iterations}. Observe the subtle changes between each iteration, attributed to the influence of the objective function used to estimate $H_t$ (see \cref{eq:h_optim}).}    
    \label{fig:hess_visual} 
    \vspace{-0.1em}
\end{figure*}

\begin{figure*}[t] 
    \centering 
    \includegraphics[width=1.0\textwidth]{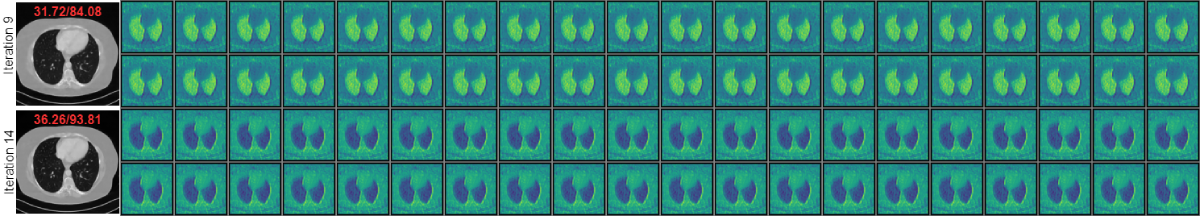}
    \vspace{-0.5cm}
    \caption{\textbf{Inverse Hessian approximation rows visualization}. We present the first $40$ rows for the $9^{th}$ and $14^{th}$ inverse Hessian approximations on the first and second lines, respectively. Each row is of size $64^2$, reshaped into a $64 \times 64$ image. The corresponding image reconstructions are shown on the left, along with PSNR (dB) and SSIM (\%) values at the top.}
    \label{fig:hess_visual_rows} 
    \vspace{-0.5cm}
\end{figure*}

%% file: apx/fig/hess_req.tex
\begin{figure}[htbp]
    \centering
    \captionsetup[subfigure]{labelformat=empty}
    \subfloat[]{
    \includegraphics[width=0.45\linewidth]{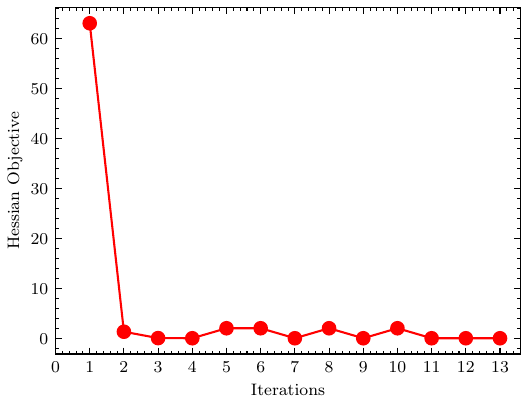}
    \label{subfig:hess_object}
    }
    \subfloat[]{
    \includegraphics[width=0.45\linewidth]{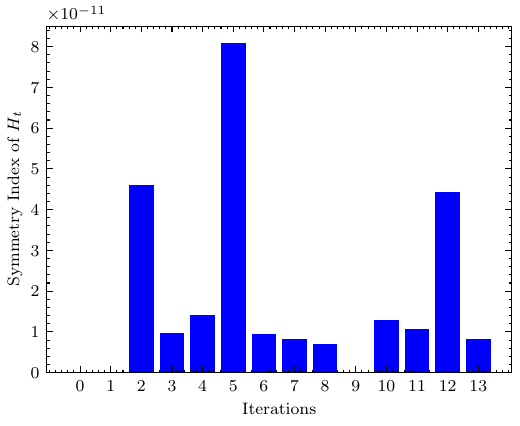}
    \label{subfig:hess_sym}
    }
    \vspace{-0.5cm}
    \caption{\textbf{Inverse Hessian matrix approximation algorithm}. Verification of requirements over the iterations. Left: Objective function value in \cref{eq:h_optim}; Right: Symmetry index of the inverse Hessian approximation refer to \cref{eq:symidx}.}
    \vspace{-0.5cm}
    \label{fig:hess_req}
\end{figure}

%% file: apx/tab/abx_ablation.tex
\begin{table}[!htb]
\centering
\resizebox{.95\linewidth}{!}{
\begin{subtable}{.5\linewidth}
  \centering
    \begin{tabular}{ccc}
        \toprule
        $p$ & {PSNR $\uparrow$} & {SSIM $\uparrow$}\\
        \midrule
        2 & 38.22 & 95.07\\
        \textbf{4} & \textbf{39.51} & \textbf{96.11}\\
        8 & 38.19 & 95.32\\
        \bottomrule
    \end{tabular}
    \caption{}
    \label{tab:reg_patch}
\end{subtable}
\begin{subtable}{.5\linewidth}
  \centering
    \begin{tabular}{ccc}
        \toprule
        $N$ & {PSNR $\uparrow$} & {SSIM $\uparrow$}\\
        \midrule
        1 & 37.64 & 94.79\\
        \textbf{2} & \textbf{39.51} & \textbf{96.11}\\
        3 & 38.47 & 95.51\\
        4 & 38.17 & 95.40\\
        \bottomrule
    \end{tabular}
    \caption{}
    \label{tab:reg_depth}
\end{subtable}
}
\caption{\textbf{Ablation of \texttt{\regterm}}. (a) $p$ is the patch size; (b) $N$ is the number of stacked Mixer layers. The best performance is attained using $p=4$ and $N=2$.}
\label{tab:ablation_results}
\end{table}

%% file: apx/fig/it_fig.tex
\begin{figure*}
    \centering 
    \includegraphics[width=1.0\textwidth]{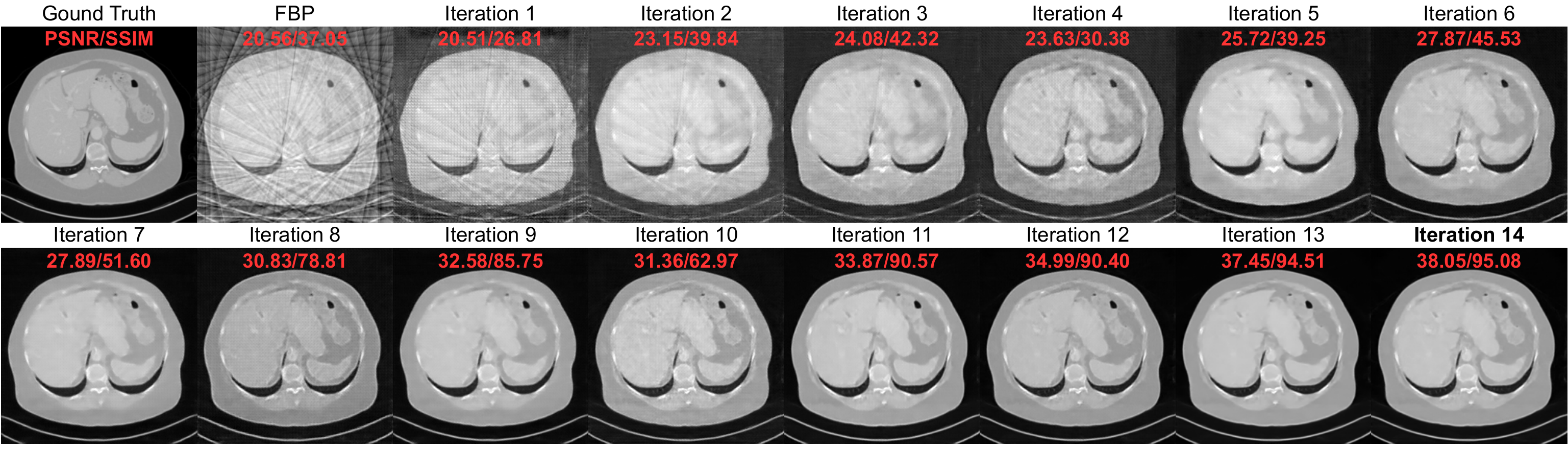}
    \vspace{-0.5cm}
    \caption{\textbf{\methodname's intermediate reconstructions} using AAPM with $(n_v = 32, N_1)$. Display window is set to $\left[-1000, 800\right]$ HU.}    
    \label{fig:it_fig} 
    \vspace{-0.1em}
\end{figure*}

%% file: apx/fig/reconstruction.tex
\begin{figure}[htp!]
    \captionsetup[subfigure]{labelformat=empty}
    \subfloat[LEARN]{
      \includegraphics[clip,width=\columnwidth]{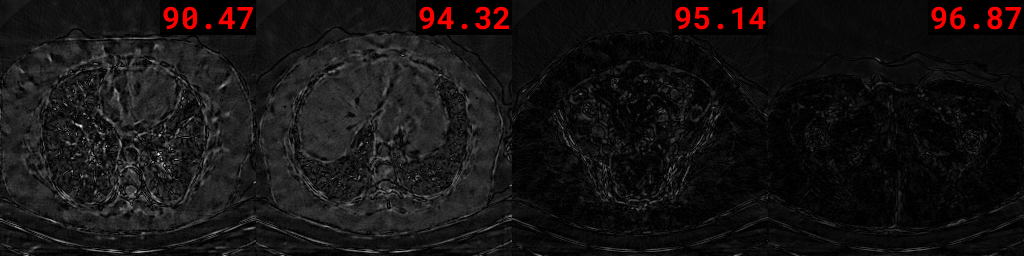}
    }

    \subfloat[RegFormer]{
      \includegraphics[clip,width=\columnwidth]{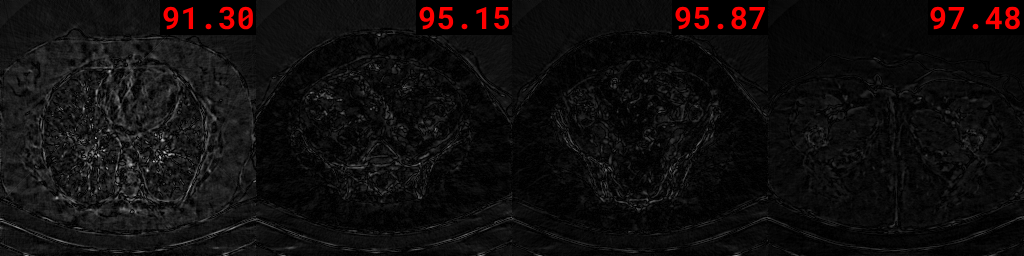}
    }

    \subfloat[\textbf{\methodname}]{
      \includegraphics[clip,width=\columnwidth]{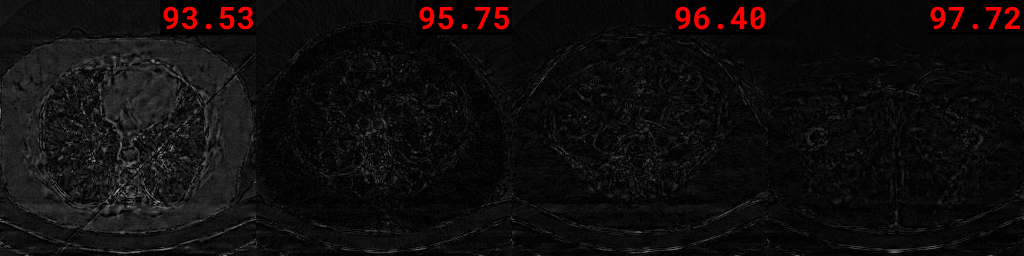}
    }
    \vspace{-0.25cm}
    \caption{
        \textbf{Reconstruction errors} with LEARN, RegFormer, and \methodname using $(n_v = 32, N_0)$. Images are ordered left to right by SSIM, with the first column showing the worst reconstruction among $214$ patient images. The second and third columns represent the $1/3$ and $2/3$ percentiles, respectively, and the last column corresponds to the best reconstruction with the highest SSIM.
    }
    \label{apx:fig_reconstruction}
\end{figure}

%% file: apx/tab/ood_patient.tex
\begin{table}[htpb]
\centering
\resizebox{.45\textwidth}{!}{
\begin{tabular}{lrrrrrr}
    \toprule
    \multirow{2}{*}{Method} & \multicolumn{2}{c}{\textbf{$n_v  =  32$}} & \multicolumn{2}{c}{\textbf{$n_v  =  64$}} & \multicolumn{2}{c}{\textbf{$n_v  =  128$}} \\
    \cmidrule(lr){2-3}\cmidrule(lr){4-5}\cmidrule(lr){6-7}
     & SSIM $\uparrow$ & PSNR $\uparrow$ & SSIM $\uparrow$ & PSNR $\uparrow$ & SSIM $\uparrow$ & PSNR $\uparrow$ \\
    \midrule
    FBP & 72.88 & 18.97 & 83.42 & 22.13 & \underline{91.75} & 24.85 \\
    FBPConvNet~\cite{fbpconvnet} & 63.91 & 20.94 & 73.02 & 24.12 & 80.60 & 25.74 \\
    DuDoTrans~\cite{dudotrans} & 60.51 & 19.09 & 79.75 & 25.00 & 85.65 & 27.23 \\
    Learned PD~\cite{learnedpd} & 67.99 & 21.92 & 83.79 & 25.51 & 85.42 & 25.86 \\
    LEARN~\cite{learn} & \underline{79.70} & \underline{24.46} & \underline{84.44} & \underline{26.74} & 88.16 & 26.20 \\
    RegFormer~\cite{regformer} & 72.45 & 23.69 & 77.33 & 25.46 & 84.99 & \underline{28.22} \\
    \midrule
    QN-Mixer~\textcolor{trolleygrey}{(ours)} & \textbf{86.17} & \textbf{25.95} & \textbf{94.56} & \textbf{30.95} & \textbf{97.04} & \textbf{33.98} \\
    \bottomrule
\end{tabular}
}
\caption{\textbf{Quantitative results of the reconstruction of the cropped OOD circle.} \textbf{Bold}: Best, \underline{under}: second best.}
\label{apx:ood_patient}
\end{table}

%% file: apx/fig/ood_fig.tex
\begin{figure*}[ht]
    \centering
    \begin{subfigure}[b]{0.29\linewidth}
        \centering
        \includegraphics[width=\linewidth]{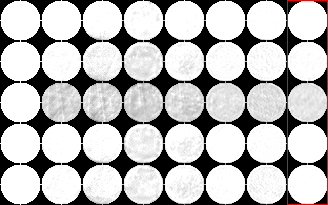}
        \caption{$n_v = 32, N_0$}
        \label{fig:fig1a}
    \end{subfigure}
    \quad
    \begin{subfigure}[b]{0.29\linewidth}
        \centering
        \includegraphics[width=\linewidth]{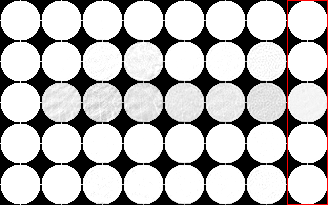}
        \caption{$n_v = 64, N_0$}
        \label{fig:fig1b}
    \end{subfigure}
    \quad
    \begin{subfigure}[b]{0.29\linewidth}
        \centering
        \includegraphics[width=\linewidth]{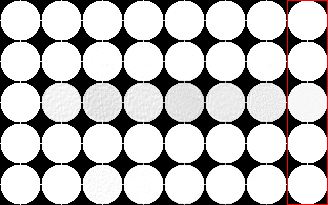}
        \caption{$n_v = 128, N_0$} 
        \label{fig:fig1c}
    \end{subfigure}
    \vspace{-0.25cm}
    \caption{
        \textbf{Visualization of 5 samples of the OOD circle texture reconstruction.}
        From each figure and from left to right, we show the ground truth,
        FBP, FBPConvNet, DuDoTrans, Learned PD, LEARN, RegFormer and \methodname.
    }
    \label{apx:ood_fig}
\end{figure*}

%% file: apx/tab/ood_hard.tex
\begin{table}[htpb]
\centering
\resizebox{.45\textwidth}{!}{
\begin{tabular}{l|cccccc}
\toprule
\textbf{Train} & \multicolumn{2}{c}{AAPM \scriptsize{$n_v{=}32$}} & \multicolumn{4}{c}{AAPM \scriptsize{$n_v{=}128$}} \\
\cmidrule(lr){2-3}
\cmidrule(lr){4-7}
\textbf{Test} & \multicolumn{2}{c}{DeepLesion \scriptsize{$n_v{=}32$}} & \multicolumn{2}{c}{AAPM \scriptsize{$n_v{=}32$}} & \multicolumn{2}{c}{AAPM \scriptsize{$n_v{=}64$}}\\
\cmidrule(lr){2-3}
\cmidrule(lr){4-5}
\cmidrule(lr){6-7}
Method & {PSNR $\uparrow$} & SSIM $\uparrow$ & PSNR $\uparrow$ & SSIM $\uparrow$ & PSNR $\uparrow$ & SSIM $\uparrow$\\
\midrule
DuDoTrans & 28.69 & 73.77 & 25.60 & 51.01 & 32.38 & 82.35 \\
LEARN & 37.21 & 93.46 & 28.20 & 66.10 & 34.40 & 87.21\\
RegFormer & \underline{37.97} & \underline{94.01} & \underline{29.83} & \underline{72.53} & \textbf{37.78} & \textbf{92.64}\\
QN-Mixer~\textcolor{trolleygrey}{(ours)} & \textbf{38.28} & \textbf{94.66} & \textbf{30.43} & \textbf{80.03} & \underline{37.16} & \underline{91.81}\\
\bottomrule
\end{tabular}
}
\caption{\textbf{Quantitative results of the reconstruction performance of OOD cases: anatomy and geometry} \textbf{Bold}: Best, \underline{under}: second best.}
\label{apx:ood_hard}
\end{table}

%% file: apx/fig/nps.tex
\begin{figure*}[ht]
    \centering
    \begin{subfigure}[b]{0.29\linewidth}
        \centering
        \includegraphics[width=\linewidth]{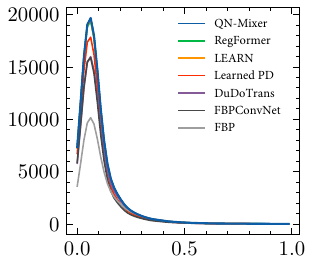}
        \caption{$n_v = 32, N_1$}
        \label{fig:nps_32}
    \end{subfigure}
    \quad
    \begin{subfigure}[b]{0.29\linewidth}
        \centering
        \includegraphics[width=\linewidth]{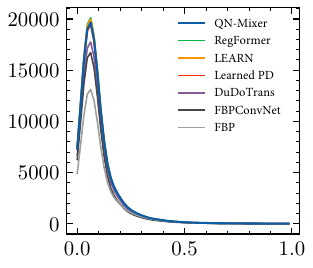}
        \caption{$n_v = 64, N_1$}
        \label{fig:nps_64}
    \end{subfigure}
    \quad
    \begin{subfigure}[b]{0.29\linewidth}
        \centering
        \includegraphics[width=\linewidth]{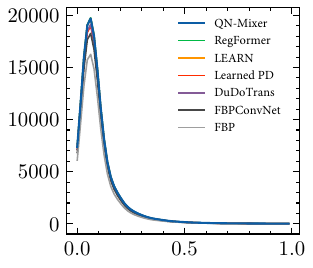}
        \caption{$n_v = 128, N_1$} 
        \label{fig:nps_128}
    \end{subfigure}
    \vspace{-0.25cm}
    \caption{
        \textbf{Noise Power Spectrum (NPS) Analysis} in comparison to state-of-the-art methods. The x-axis represents normalized frequency in cycles per pixel ($\text{px}^{-1}$), and the y-axis represents noise power spectrum ($\text{HU}^2 \text{px}^2$). Display windows are configured as $\left[-1000, 800\right]$ HU. Mean and peak frequencies are intricately linked to noise texture, with finer textures correlating to higher mean and peak frequencies in the NPS. Our method exhibits the highest peak frequencies, indicating that our reconstructed images feature the most refined noise texture among all compared methods.
    }    
    \label{apx:nps}
\end{figure*}

%% file: apx/fig/nps_roi.tex
\begin{figure}
    \centering 
    \includegraphics[width=0.5\columnwidth]{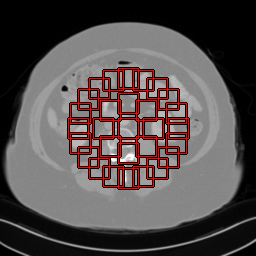}
    \caption{
        \textbf{ROIs for NPS Analysis:} Red squares denote $20 \times 20$ pixel ROIs distributed evenly across two circular regions. The first circle (radius $25$) holds $8$ ROIs, and the second circle (radius $50$) has $20$ ROIs. Both circles, centered at the image center, include a total of $29$ ROIs per image. This standard positioning in the CT community underscores the clinical diagnostic importance of the image center.
    }
    \label{fig:nps_roi} 
\end{figure}

%% file: apx/alg/qn_mixer.tex
\newcommand{\hlbox}[1]{
  \fboxsep=1.2pt\hspace*{-\fboxsep}\colorbox{blue!10}{\detokenize{#1}}
}
\begin{algorithm}[b]
\caption{Minimal \methodname pseudo-code}\label{apx:qn_mixer}
\lstset{style=mocov3}
\vspace{-3pt}
\begin{lstlisting}[
    language=python,
    escapechar=@,
    label=code:qn_mixer]
class GradientFunction(nn.Module):
  def __init__(self, regularizer):
    @\hlbox{self.regularizer}@ = regularizer
    @\hlbox{self.lambda}@ = nn.Parameter(torch.zeros(1))

  def forward(self, physics, y, x):
    y_t = physics.forward_operator(x)
    # Compute the regularization term
    reg_x = self.regularizer(x)
    # Compute the data fidelity term
    y_dft = y_t - y
    # Compute the backprojection
    x_dft = physics.backward_operator(y_dft)
    @\hlbox{g = self.lambda * x_dft + reg_x}@
    return g
  
class QN_Iteration(nn.Module):
  def __init__(self, gradient_function):
    @\hlbox{self.gradient}@ = gradient_function

  def @\hlbox{latent_bfgs}@(self, h, s_t, z_t):
    I = torch.eye(len(s_t))
    rho_t = 1. / torch.dot(z_t, s_t)
    u_t = I - torch.outer(s_t, z_t) * rho_t
    d_t = I - torch.outer(z_t, s_t) * rho_t
    return (torch.matmul(u_t, torch.matmul(h, d_t)) 
                + (torch.outer(s_t, s_t) * rho_t))

  def forward(self, physics, encoder, decoder,
                    y, x, h, r, is_last):
    # Compute latent direction s_t
    @\hlbox{s_t = -torch.matmul(h, r)}@
    @\hlbox{d = decoder(s_t)}@

    # Update the reconstruction
    @\hlbox{x = x + d}@
    # Return x if it is the last iteration
    if is_last:
        return x, h, r
    else:
        @\hlbox{r_p = encoder(self.gradient(physics, y, x))}@
        z_t = r_p - r
        with torch.no_grad():
            h = self.latent_bfgs(h, s_t, z_t)
        return x, h, r_p
\end{lstlisting}
\vspace{-5pt}
\end{algorithm}

%% file: apx/tab/aapm_splits.tex
\begin{table}[htpb]
\centering
\resizebox{.45\textwidth}{!}{
\begin{tabular}{lc@{\hskip 15pt}ccccccccc}
    \toprule
    Patient ID & L067 & L109 & L143 & L192 & L286 & L291 & L096 & L506 & L333 & L310 \\
    \toprule
    \#slices & 224 & 128 & 234 & 240 & 210 & 343 & 330 & 211 & 244 & 214 \\
    \toprule
    Training & \cmark & \cmark & \cmark & \cmark & \cmark & \cmark & \cmark & \cmark & \grayxmark & \grayxmark \\
    \toprule
    Validation & \grayxmark & \grayxmark & \grayxmark & \grayxmark & \grayxmark & \grayxmark & \grayxmark & \grayxmark & \cmark & \grayxmark \\
    \toprule
    Testing & \grayxmark & \grayxmark & \grayxmark & \grayxmark & \grayxmark & \grayxmark & \grayxmark & \grayxmark & \grayxmark & \cmark \\
    \bottomrule
\end{tabular}
}
\caption{\textbf{AAPM dataset split specification}. The validation set comprises images from patient L333, and testing utilizes images from patient L310. The images from the remaining patients have been designated for training purposes.}    
\label{tab:aapm_splits}
\end{table}

%% file: apx/tab/abx_arch.tex
\begin{table}[b]
\centering
\resizebox{\linewidth}{!}{
    \begin{tabular}{l@{\hskip 15pt}ccc}
        \toprule
        \textbf{Stage} & \textbf{Layers} & \textbf{\#Param (k)} & \textbf{Output size} \\
        \midrule
        \midrule
        Input & - & - & $1\times 256 \times 256$ \\ 
        \midrule
        InceptionBlock-1 & 
            \begin{tabular}[c]{@{}l@{}}
                convblock1: 
                    \begin{tabular}[c]{@{}l@{}}
                        conv1-1: K1C16S1P0 \\
                        prelu1-1 \\
                    \end{tabular} \\          
                \midrule
                convblock2: 
                    \begin{tabular}[c]{@{}l@{}}
                        conv2-1: K1C16S1P0 \\
                        prelu2-1 \\
                        conv2-2: K3C32S1P1 \\
                        prelu2-2 \\
                    \end{tabular} \\                 
                \midrule
                convblock3: 
                    \begin{tabular}[c]{@{}l@{}}
                        conv3-1: K1C16S1P0 \\
                        prelu3-1 \\
                        conv3-2: K5C32S1P2 \\
                        prelu3-2 \\
                    \end{tabular} \\
                \midrule
                convblock4: 
                    \begin{tabular}[c]{@{}l@{}}
                        maxpool4-1: K3S1P1 \\
                        conv4-1: K1C16S1P0 \\
                        prelu4-1 \\
                    \end{tabular} \\                    
            \end{tabular}
        & 17.6 & $96\times 256\times 256$ \\ 
        \midrule
        PatchEmbed-2 & 
            \begin{tabular}[c]{@{}l@{}}
                conv2-1: K4C96S4P0 \\
                rearrange2-1: bchw $\rightarrow$ bhwc \\
            \end{tabular}
        & $145.5$ & $96\times 64\times 64$  \\ 
        \midrule
        MixerLayer-3 $\times (N = 2)$ & 
            \begin{tabular}[c]{@{}l@{}}
                layernorm3-1: D96 \\
                rearrange3-1: bhwc $\rightarrow$ bcwh \\
                \midrule
                heightmlp3-1: 
                    \begin{tabular}[c]{@{}l@{}}
                        linear3-1: D64O256 \\
                        gelu3-1 \\
                        linear3-2: D256O64 \\                        
                    \end{tabular} \\          
                \midrule
                rearrange3-2: bcwh $\rightarrow$ bchw \\
                \midrule
                widthmlp3-1: 
                    \begin{tabular}[c]{@{}l@{}}
                        linear3-3: D64O256 \\
                        gelu3-2 \\
                        linear3-4: D256O64 \\                        
                    \end{tabular} \\
                \midrule
                rearrange3-3: bchw $\rightarrow$ bhwc \\
                layernorm3-2: D96 \\
                \midrule
                channelmlp3-1: 
                    \begin{tabular}[c]{@{}l@{}}
                        linear3-5: D96O384 \\
                        gelu3-3 \\
                        linear3-6: D384O96 \\                        
                    \end{tabular} \\                
            \end{tabular}        
        & $140.8 \times 2$ & $96\times 64\times 64$ \\ 
        \midrule
        PatchExpand-4 & 
            \begin{tabular}[c]{@{}l@{}}
                linear4-1: D96O1536 \\
                layernorm4-1: D96 \\
                conv4-1: K1C1S1P0 \\
            \end{tabular}        
        & $147.7$ & $1\times 256\times 256$  \\
        \bottomrule
    \end{tabular}
}
\caption{\textbf{\regterm architecture}. \texttt{K-C-S-P} represents the kernel, channel, stride, and padding configuration of CNNs, while \texttt{D-O} indicates the input and output dimensions of linear layers.}
\label{tab:abx_arch}
\end{table}

%% file: apx/alg/circle.tex
\begin{algorithm}[htpb]
\caption{\texttt{add\_circle\_ood} pseudo-code.}\label{apx:circle}
\lstset{style=mocov3}
\vspace{-3pt}
\begin{lstlisting}[
    language=python,
    escapechar=@,
    label=code:circle]
def add_circle_ood(img, value=1):
  h, w = img.shape[::-1][:2]
  radius = np.random.randint(5, 20)
  c_x = np.random.randint(radius, w-radius)
  c_y = np.random.randint(radius, h-radius)
  center = (c_x, c_y)
    
  Y, X = np.ogrid[:h, :w]
  dist_x = (X - center[0])**2
  dist_y = (Y - center[1])**2
  dist_from_center = np.sqrt(dist_x + dist_y)
  mask = dist_from_center <= radius
  img[0, mask] = value
  return img
\end{lstlisting}
\vspace{-5pt}
\end{algorithm}

%% file: apx/tab/notations.tex
\begin{table*}
    \definecolor{Gray}{gray}{0.9}
    \centering
    \resizebox{\textwidth}{!}{
    \begin{tabular}{cccl}

        \toprule
        Notation & Shape & Value(s) & Description \\
        \midrule
        \midrule
        $n_v$ & $\mathbb{N}^{*}$ & $\{32, 64, 128\}$ & The number of projection views \\
        $n_d$ & $\mathbb{N}^{*}$ & 512 & The number of projection detectors \\
        $h$ & $\mathbb{N}^{*}$ & $256$ & Height of the image \\
        $w$ & $\mathbb{N}^{*}$ & $256$ & Width of the image \\
        $c$ & $\mathbb{N}^{*}$ & $1$ & Channels of the image \\
        $l_h$ & $\mathbb{N}^{*}$ & $64$ & Latent height\\
        $l_w$ & $\mathbb{N}^{*}$ & $64$ & Latent width\\
        $m=n_v \times n_d$ & $\mathbb{N}^{*}$ & $n_v \times 512$ & Data (sinogram) size \\
        $n=h \times w$ & $\mathbb{N}^{*}$ & $256 \times 256$ & Image size \\
        $(l_h \cdot l_w) \times (l_h \cdot l_w)$ & $\R$ & $(64 \cdot 64) \times (64 \cdot 64)$ & Size of the latent BFGS optimization variable i.e. $\mH$\\
        $T$ & $\mathbb{N}^{*}$ & 14 & Number of iterations of our method \\
        $t$ & $\mathbb{N}$ & - & Iteration of the loop in the algorithm \\
        $\vy$ & $n_v \times n_d$ & - & Sparse sinogram \\
        $\mA$ & $\R^{n \times m}$ & - & The forward model (i.e. discrete Radon transform) \\
        $\mA^{\dagger}$ & $\R^{m \times n}$ & - & The pseudo-inverse of $\mA$ \\
        $\vx_0$ & $\R^{h \times w \times c}$ & $\mA^{\dagger}\vy$ & Initial reconstruction \\
        $\lambda_t$ & $\R$ & - & Regularization weight at step $t$ \\
        $\alpha_t$ & $\R$ & - & Step size (i.e. search step) \\
        $\vx_t$ & $\R^{h \times w \times c}$ & - & Reconstructed image at iteration $t$ \\
        $\nabla_\vx J(\vx_t)$ & $\R^{h \times w \times c}$ & - & Gradient value at iteration $t$ \\
        $\mH_t$ & $\R^{(l_h \cdot l_w) \times (l_h \cdot l_w)}$ & - & Approximation of the inverse Hessian matrix at iteration $t$ \\
        $\mI^{n \times n}$ & $\mathbb{N}^{n \times n}$ & - & Identity matrix of size $n\times n$ \\
        $\vf_t$ & $\R^{h \times w \times d}$ & - & Feature map after the Inception block at iteration $t$ \\
        $\ve_t$ & $\R^{\frac{h}{p} \times \frac{w}{p} \times d}$ & - & MLP-Mixer embeddings \\
        $d$ & $\R$ & $96$ & Depth of features \\
        $p$ & $\mathbb{N}^{*}$ & $4$ & Stride and kernel size in the patchification Conv 2D net \\
        $N$ & $\mathbb{N}^{*}$ & $2$ & Number of stacked Mixer layers \\
        $\regfunc{G}(\cdot)$ & - & - & Learned gradient of the regularization term (i.e. the Incept-Mixer model) \\
        $\regfunc{G}(\vx_{t})$ & $\R^{h \times w \times c}$ & - & Regularization term at step $t$ \\        
        $\mathcal{E}(\cdot)$ & - & - & The gradient encoder \\
        $\mathcal{D}(\cdot)$ & - & - & The direction decoder \\
        $k$ & $\mathbb{N}^{*}$ & $\{2, 3, 4, 5\}$ & Number of Downsampling stacks in the encoder \\
        $\vf_{\mathcal{E}}=2^{k}$ & $\mathbb{N}^{*}$ & $\{4, 8, 16, 32\}$ & Downsampling factor of the gradient in the encoder \\
        $w_l={w}/{\vf_{\mathcal{E}}}$ & $\mathbb{N}^{*}$ & $\{64, 32, 16, 8\}$ & Number of columns of the down-sampled gradient \\
        $h_l={h}/{\vf_{\mathcal{E}}}$ & $\mathbb{N}^{*}$ & $\{64, 32, 16, 8\}$ & Number of rows of the down-sampled gradient \\
        $\vr_t = \mathcal{E}(\nabla_{\vx} J(\vx_t))$ & $\R^{l_h \cdot l_w}$ & - & Latent representation of the gradient \\
        $s_t=-\mH_t \vr_t$ & $\R^{l_h \cdot l_w}$ & - & Direction in the latent space \\
        $\rho_t = {(\vz_t^\mathsf{T} \vs_t)}^{-1}$ & $\R^{l_h \cdot l_w}$ & - & BFGS divider variable \\
        $N_0$ & - & - & Zero noise added to the sinogram \\
        $N_1$ & - & - & 5\% Gaussian noise, $1 \times 10^6$ intensity Poisson noise \\
        $N_2$ & - & - & 5\% Gaussian noise, $5 \times 10^5$ intensity Poisson noise \\
        \bottomrule
    \end{tabular}
    }
    \caption{\textbf{Lookup table of notations and hyperparameters} used in the paper.}
    \label{apx:notations}
\end{table*}

%% file: main.bbl
\begin{thebibliography}{59}
\providecommand{\natexlab}[1]{#1}
\providecommand{\url}[1]{\texttt{#1}}
\expandafter\ifx\csname urlstyle\endcsname\relax
  \providecommand{\doi}[1]{doi: #1}\else
  \providecommand{\doi}{doi: \begingroup \urlstyle{rm}\Url}\fi

\bibitem[Adler and {\"O}ktem(2018)]{learnedpd}
Jonas Adler and Ozan {\"O}ktem.
\newblock Learned primal-dual reconstruction.
\newblock \emph{IEEE TMI}, 37:\penalty0 1322--1332, 2018.

\bibitem[Arie and Avinash(1984)]{sart}
Andersen Arie and Kak Avinash.
\newblock Simultaneous algebraic reconstruction technique ({SART}): a superior
  implementation of the art algorithm.
\newblock \emph{Ultrasonic imaging}, 6:\penalty0 81--94, 1984.

\bibitem[Ashish et~al.(2017)Ashish, Noam, Niki, Jakob, Llion, N, Lukasz, and
  Illia]{attention}
Vaswani Ashish, Shazeer Noam, Parmar Niki, Uszkoreit Jakob, Jones Llion,
  Gomez~Aidan N, Kaiser Lukasz, and Polosukhin Illia.
\newblock Attention is all you need.
\newblock In \emph{NeurIPS}, 2017.

\bibitem[Avinash and Malcolm(2001)]{art}
Kak Avinash and Slaney Malcolm.
\newblock \emph{{Principles of Computerized Tomographic Imaging}}.
\newblock Society for Industrial and Applied Mathematics, 2001.

\bibitem[Chen et~al.(2021)Chen, Tachella, and Davies]{ei}
Dongdong Chen, Julián Tachella, and Mike~E. Davies.
\newblock Equivariant imaging: Learning beyond the range space.
\newblock In \emph{ICCV}, pages 4359--4368, 2021.

\bibitem[Chen et~al.(2017)Chen, Zhang, Kalra, Lin, Chen, Liao, Zhou, and
  Wang]{redcnn}
Hu Chen, Yi Zhang, Mannudeep~K. Kalra, Feng Lin, Yang Chen, Peixi Liao, Jiliu
  Zhou, and Ge Wang.
\newblock Low-dose {CT} with a residual encoder-decoder convolutional neural
  network.
\newblock \emph{IEEE TMI}, 36:\penalty0 2524--2535, 2017.

\bibitem[Chen et~al.(2018)Chen, Zhang, Chen, Zhang, Zhang, Sun, Lv, Liao, Zhou,
  and Wang]{learn}
Hu Chen, Yi Zhang, Yunjin Chen, Junfeng Zhang, Weihua Zhang, Huaiqiang Sun,
  Yang Lv, Peixi Liao, Jiliu Zhou, and Ge Wang.
\newblock {LEARN}: Learned experts' assessment-based reconstruction network for
  sparse-data {CT}.
\newblock \emph{IEEE TMI}, 37:\penalty0 1333--1347, 2018.

\bibitem[Cheng et~al.(2020)Cheng, Wang, Li, and Duan]{cheng2020}
Weilin Cheng, Yu Wang, Hongwei Li, and Yuping Duan.
\newblock Learned full-sampling reconstruction from incomplete data.
\newblock \emph{IEEE TCI}, pages 945--957, 2020.

\bibitem[Chengbin et~al.(1993)Chengbin, L., and Toks{\"o}z]{tikhonov}
Peng Chengbin, Rodi~William L., and M.~Nafi Toks{\"o}z.
\newblock \emph{A Tikhonov Regularization Method for Image Reconstruction},
  pages 153--164.
\newblock Springer US, 1993.

\bibitem[Davidon(1991)]{qnbfgs}
William~C. Davidon.
\newblock Variable metric method for minimization.
\newblock \emph{SIAM Journal on Optimization}, 1:\penalty0 1--17, 1991.

\bibitem[Dianlin et~al.(2022)Dianlin, Yikun, Jin, Shouhua, and Yang]{dior}
Hu Dianlin, Zhang Yikun, Liu Jin, Luo Shouhua, and Chen Yang.
\newblock {DIOR}: Deep iterative optimization-based residual-learning for
  limited-angle {CT} reconstruction.
\newblock \emph{IEEE TMI}, pages 1778--1790, 2022.

\bibitem[Fabian et~al.(2022)Fabian, Tinaz, and Soltanolkotabi]{humusnet}
Zalan Fabian, Berk Tinaz, and Mahdi Soltanolkotabi.
\newblock {HUMUS}-{N}et: Hybrid unrolled multi-scale network architecture for
  accelerated {MRI} reconstruction.
\newblock In \emph{NeurIPS}, 2022.

\bibitem[Fletcher(1987)]{bfgs}
Roger Fletcher.
\newblock \emph{Practical Methods of Optimization}.
\newblock John Wiley \& Sons, New York, NY, USA, 1987.

\bibitem[Gartner et~al.(2023)Gartner, Metz, Andriluka, Freeman, and
  Sminchisescu]{optimus}
Erik Gartner, Luke Metz, Mykhaylo Andriluka, C.~Daniel Freeman, and Cristian
  Sminchisescu.
\newblock Transformer-based {L}earned {O}ptimization.
\newblock In \emph{CVPR}, pages 11970--11979, 2023.

\bibitem[Ghani and Karl(2018)]{sinogramcom}
Muhammad~Usman Ghani and W.~Clem Karl.
\newblock Deep learning-based sinogram completion for low-dose {CT}.
\newblock In \emph{2018 IEEE 13th Image, Video, and Multidimensional Signal
  Processing Workshop}, pages 1--5, 2018.

\bibitem[Harshit et~al.(2018)Harshit, Hwan, Q., T., and Michael]{cnngrad}
Gupta Harshit, Jin~Kyong Hwan, Nguyen~Ha Q., McCann~Michael T., and Unser
  Michael.
\newblock {CNN}-based projected gradient descent for consistent {CT} image
  reconstruction.
\newblock \emph{IEEE TMI}, 37:\penalty0 1440--1453, 2018.

\bibitem[Hendriksen et~al.(2020)Hendriksen, Pelt, and Batenburg]{n2inv}
Allard~Adriaan Hendriksen, Daniël~Maria Pelt, and K.~Joost Batenburg.
\newblock Noise2inverse: Self-supervised deep convolutional denoising for
  tomography.
\newblock \emph{IEEE TCI}, pages 1320--1335, 2020.

\bibitem[Hu et~al.(2020)Hu, Liu, Lv, Zhao, Zhang, Quan, Feng, Chen, and
  Luo]{hdnet}
Dianlin Hu, Jin Liu, Tianling Lv, Qianlong Zhao, Yikun Zhang, Guotao Quan, Juan
  Feng, Yang Chen, and Limin Luo.
\newblock Hybrid-domain neural network processing for sparse-view {CT}
  reconstruction.
\newblock \emph{IEEE TRPMS}, 5:\penalty0 88--98, 2020.

\bibitem[Jin et~al.(2017)Jin, McCann, Froustey, and Unser]{fbpconvnet}
Kyong~Hwan Jin, Michael~T. McCann, Emmanuel Froustey, and Michael Unser.
\newblock Deep convolutional neural network for inverse problems in imaging.
\newblock \emph{IEEE TIP}, 26:\penalty0 4509--4522, 2017.

\bibitem[Jinxi et~al.(2021)Jinxi, Yonggui, and Yunjie]{fistanet}
Xiang Jinxi, Dong Yonggui, and Yang Yunjie.
\newblock {FISTA}-{N}et: Learning a fast iterative shrinkage thresholding
  network for inverse problems in imaging.
\newblock \emph{IEEE TMI}, 40:\penalty0 1329--1339, 2021.

\bibitem[Jorge and J(2006)]{qn}
Nocedal Jorge and Wright~Stephen J.
\newblock Quasi-{N}ewton methods.
\newblock \emph{Numerical optimization}, 75:\penalty0 135--163, 2006.

\bibitem[Kawata and Nalcioglu(1985)]{cgls}
Satoshi Kawata and Orhan Nalcioglu.
\newblock Constrained iterative reconstruction by the conjugate gradient
  method.
\newblock \emph{IEEE TMI}, 4:\penalty0 65--71, 1985.

\bibitem[Kobler et~al.(2020)Kobler, Effland, Kunisch, and
  Pock]{kobler2020total}
Erich Kobler, Alexander Effland, Karl Kunisch, and Thomas Pock.
\newblock Total deep variation for linear inverse problems.
\newblock In \emph{CVPR}, pages 7546--7555, 2020.

\bibitem[Lee et~al.(2018)Lee, Lee, Kim, Cho, and Cho]{ssnet}
Hoyeon Lee, Jongha Lee, Hyeongseok Kim, Byungchul Cho, and Seungryong Cho.
\newblock Deep-neural-network-based sinogram synthesis for sparse-view {CT}
  image reconstruction.
\newblock \emph{IEEE TRPMS}, 3:\penalty0 109--119, 2018.

\bibitem[Lee et~al.(2020)Lee, Kim, and Kim]{leemwcnn}
Minjae Lee, Hyemi Kim, and Hee-Joung Kim.
\newblock Sparse-view {CT} reconstruction based on multi-level wavelet
  convolution neural network.
\newblock \emph{Physica Medica}, 80:\penalty0 352--362, 2020.

\bibitem[Li et~al.(2020)Li, Hsu, Xie, Cong, and Gao]{SACNN}
Meng Li, William Hsu, Xiaodong Xie, Jason Cong, and Wen Gao.
\newblock Sacnn: Self-attention convolutional neural network for low-dose ct
  denoising with self-supervised perceptual loss network.
\newblock \emph{IEEE TMI}, pages 2289--2301, 2020.

\bibitem[Li et~al.(2022)Li, Li, Wang, Li, Zhao, Qiang, and
  Wang]{ddptransformer}
Runrui Li, Qing Li, Hexi Wang, Saize Li, Juanjuan Zhao, Yan Qiang, and Long
  Wang.
\newblock {DDPTransformer}: Dual-domain with parallel transformer network for
  sparse view {CT} image reconstruction.
\newblock \emph{IEEE TCI}, pages 1--15, 2022.

\bibitem[Li et~al.(2023)Li, Ma, Chen, Zhang, and Shan]{gloredi}
Zilong Li, Chenglong Ma, Jie Chen, Junping Zhang, and Hongming Shan.
\newblock Learning to distill global representation for sparse-view ct.
\newblock In \emph{ICCV}, pages 21196--21207, 2023.

\bibitem[Lin et~al.(2019)Lin, Liao, Peng, Sun, Zhang, Luo, Chellappa, and
  Kevin]{dudonet}
Wei-An Lin, Cheng Liao, Haofu Peng, Xiaohang Sun, Jingdan Zhang, Jiebo Luo,
  Rama Chellappa, and Zhou~S. Kevin.
\newblock {DuDoNet}: Dual domain network for {CT} metal artifact reduction.
\newblock In \emph{CVPR}, pages 10512--10521, 2019.

\bibitem[Liu and Luo(2022)]{saddleqn}
Chengchang Liu and Luo Luo.
\newblock Quasi-newton methods for saddle point problems.
\newblock In \emph{NeurIPS}, 2022.

\bibitem[Liu et~al.(2021)Liu, Lin, Cao, Hu, Wei, Zhang, and Guo]{swint}
Ze Liu, Yutong Lin, Yue Cao, Han Hu, Yixuan Wei, Zheng Zhang, and Stephen
  Lin~Baining Guo.
\newblock Swin transformer: Hierarchical vision transformer using shifted
  windows.
\newblock In \emph{ICCV}, pages 9992--10002, 2021.

\bibitem[Loshchilov and Hutter(2017)]{adamw}
Ilya Loshchilov and Frank Hutter.
\newblock {Decoupled Weight Decay Regularization}.
\newblock \emph{arXiv preprint arXiv:1711.05101}, 2017.

\bibitem[Lunz et~al.(2018)Lunz, \"{O}ktem, and Sch\"{o}nlieb]{lunz2018}
Sebastian Lunz, Ozan \"{O}ktem, and Carola-Bibiane Sch\"{o}nlieb.
\newblock Adversarial regularizers in inverse problems.
\newblock In \emph{NeurIPS}, 2018.

\bibitem[Macleod(1963)]{fbp}
Cormack~Allan Macleod.
\newblock Representation of a function by its line integrals, with some
  radiological applications.
\newblock \emph{Journal of Applied Physics}, 34:\penalty0 2722--2727, 1963.

\bibitem[McCollough(2016)]{aapm}
C. McCollough.
\newblock {TU-FG-207A-04}: Overview of the low dose {CT} grand challenge.
\newblock \emph{Medical Physics}, 43:\penalty0 3759--3760, 2016.

\bibitem[Metz et~al.(2022)Metz, Freeman, Harrison, Maheswaranathan, and
  Sohl-Dickstein]{adafactor}
Luke Metz, C.~Daniel Freeman, James Harrison, Niru Maheswaranathan, and Jascha
  Sohl-Dickstein.
\newblock Practical tradeoffs between memory, compute, and performance in
  learned optimizers.
\newblock \emph{arXiv preprint arXiv:2203.11860}, 2022.

\bibitem[Miller and Schauer(1983)]{alara}
Donald~L. Miller and David Schauer.
\newblock The alara principle in medical imaging.
\newblock \emph{Philosophy}, 44:\penalty0 595--600, 1983.

\bibitem[Mukherjee et~al.(2021)Mukherjee, Carioni, {\"O}ktem, and
  Sch{\"o}nlieb]{mukherjee2021endtoend}
Subhadip Mukherjee, Marcello Carioni, Ozan {\"O}ktem, and Carola-Bibiane
  Sch{\"o}nlieb.
\newblock End-to-end reconstruction meets data-driven regularization for
  inverse problems.
\newblock In \emph{NeurIPS}, 2021.

\bibitem[{\"O}ktem et~al.(2014){\"O}ktem, Adler, Kohr, and Team]{odl}
Ozan {\"O}ktem, Jonas Adler, Holger Kohr, and The~ODL Team.
\newblock Operator discretization library {(ODL)}, 2014.

\bibitem[Radon(1917)]{radon}
Johann Radon.
\newblock \"uber die bestimmung von funktionen durch ihre integralwerte l\"angs
  gewisser mannigfaltigkeiten.
\newblock \emph{Berichte {\"u}ber die Verhandlungen der
  K{\"o}niglich-S{\"a}chsischen Akademie der Wissenschaften zu Leipzig},
  69:\penalty0 262--277, 1917.

\bibitem[Robin et~al.(2022)Robin, Andreas, Dominik, Patrick, and Bj\"orn]{ldms}
Rombach Robin, Blattmann Andreas, Lorenz Dominik, Esser Patrick, and Ommer
  Bj\"orn.
\newblock High-resolution image synthesis with latent diffusion models.
\newblock In \emph{CVPR}, pages 10684--10695, 2022.

\bibitem[Rui et~al.(2017)Rui, Lu, Yan, and Hengyong]{svd}
Liu Rui, He Lu, Luo Yan, and Yu Hengyong.
\newblock Singular value decomposition-based {2D} image reconstruction for
  computed tomography.
\newblock \emph{Journal of X-ray science and technology}, 25:\penalty0
  113--134, 2017.

\bibitem[Shaohua et~al.(2014)Shaohua, Yan, Zhaoying, Jing, Wufan, Hengyong,
  Zhengrong, and Jianhua]{tgv}
Niu Shaohua, Gao Yan, Bian Zhaoying, Huang Jing, Chen Wufan, Yu Hengyong, Liang
  Zhengrong, and Ma Jianhua.
\newblock Sparse-view x-ray {CT} reconstruction via total generalized variation
  regularization.
\newblock \emph{PMB}, 59:\penalty0 2997--3017, 2014.

\bibitem[Sriram et~al.(2020)Sriram, Zbontar, Murrell, Defazio, Zitnick,
  Yakubova, Knoll, and Johnson]{endtoend}
Anuroop Sriram, Jure Zbontar, Tullie Murrell, Aaron Defazio, C.~Lawrence
  Zitnick, Nafissa Yakubova, Florian Knoll, and Patricia Johnson.
\newblock {E}nd-to-{E}nd variational networks for accelerated {MRI}
  reconstruction.
\newblock In \emph{MICCAI}, pages 64--73, 2020.

\bibitem[Szegedy et~al.(2015)Szegedy, Liu, Jia, Sermanet, Reed, Anguelov,
  Erhan, and Rabinovich]{inception}
Christian Szegedy, Wei Liu, Yangqing Jia, Pierre Sermanet, Scott Reed, Dragomir
  Anguelov, Dumitru Erhan, and Vincent Vanhoucke~Andrew Rabinovich.
\newblock Going deeper with convolutions.
\newblock In \emph{CVPR}, pages 1--9, 2015.

\bibitem[Tolstikhin et~al.(2021)Tolstikhin, Houlsby, Kolesnikov, Beyer, Zhai,
  Unterthiner, Yung, Steiner, Keysers, Uszkoreit, Lucic, and
  Dosovitskiy]{mlpmixer}
Ilya Tolstikhin, Neil Houlsby, Alexander Kolesnikov, Lucas Beyer, Xiaohua Zhai,
  Thomas Unterthiner, Jessica Yung, Andreas~Peter Steiner, Daniel Keysers,
  Jakob Uszkoreit, Mario Lucic, and Alexey Dosovitskiy.
\newblock {MLP}-mixer: An all-{MLP} architecture for vision.
\newblock In \emph{NeurIPS}, 2021.

\bibitem[Tsutomu and Yukio(2018)]{tv}
Gomi Tsutomu and Koibuchi Yukio.
\newblock Use of a {T}otal {V}ariation minimization iterative reconstruction
  algorithm to evaluate reduced projections during digital breast
  tomosynthesis.
\newblock \emph{BioMed Research International}, 2018:\penalty0 1--14, 2018.

\bibitem[Unal et~al.(2021)Unal, Ertas, and Yildirim]{N2S}
Mehmet~Ozan Unal, Metin Ertas, and Isa Yildirim.
\newblock Self-supervised training for low-dose ct reconstruction.
\newblock In \emph{ISBI}, pages 69--72, 2021.

\bibitem[Wang et~al.(2022)Wang, Shang, Zhang, Li, and Zhou]{dudotrans}
Ce Wang, Kun Shang, Haimiao Zhang, Qian Li, and S.~Kevin Zhou.
\newblock {DuDoTrans}: Dual-domain transformer for sparse-view {CT}
  reconstruction.
\newblock In \emph{Machine Learning for Medical Image Reconstruction}, pages
  84--94. Springer International Publishing, 2022.

\bibitem[Wang et~al.(2008)Wang, Yu, and Man]{ct_outlook}
Ge Wang, Hengyong Yu, and Bruno~De Man.
\newblock An outlook on {X}-ray {CT} research and development.
\newblock \emph{Medical Physics}, 35:\penalty0 1051--1064, 2008.

\bibitem[Wang et~al.(2019)Wang, Zeng, Wang, and Guo]{admmct}
Jiaxi Wang, Li Zeng, Chengxiang Wang, and Yumeng Guo.
\newblock {ADMM}-based deep reconstruction for limited-angle {CT}.
\newblock \emph{PMB}, 64, 2019.

\bibitem[Weiwen et~al.(2021)Weiwen, Dianlin, Chuang, Hengyong, Varut, and
  Ge]{drone}
Wu Weiwen, Hu Dianlin, Niu Chuang, Yu Hengyong, Vardhanabhuti Varut, and Wang
  Ge.
\newblock {DRONE}: Dual-domain residual-based optimization network for
  sparse-view {CT} reconstruction.
\newblock \emph{IEEE TMI}, 40:\penalty0 3002--3014, 2021.

\bibitem[Wu et~al.(2017)Wu, Kim, Fakhri, and Li]{wu2017iterative}
Dufan Wu, Kyungsang Kim, Georges~El Fakhri, and Quanzheng Li.
\newblock Iterative low-dose ct reconstruction with priors trained by
  artificial neural network.
\newblock \emph{IEEE TMI}, 36:\penalty0 2479--2486, 2017.

\bibitem[Xia et~al.(2022)Xia, Yang, Zhou, Lu, Wang, and Zhang]{regformer}
Wenjun Xia, Ziyuan Yang, Qizheng Zhou, Zexin Lu, Zhongxian Wang, and Yi Zhang.
\newblock Transformer-based iterative reconstruction model for sparse-view {CT}
  reconstruction.
\newblock In \emph{MICCAI}, 2022.

\bibitem[Yan et~al.(2018)Yan, Wang, Lu, and Summers]{deeplesion}
Ke Yan, Xiaosong Wang, Le Lu, and Ronald~M. Summers.
\newblock {DeepLesion}: Automated mining of large-scale lesion annotations and
  universal lesion detection with deep learning.
\newblock \emph{Journal of Medical Imaging}, 5:\penalty0 036501, 2018.

\bibitem[Yi et~al.(2023)Yi, Hu, Wenjun, Yang, Baodong, Yan, Huaiqiang, and
  Jiliu]{learnpp}
Zhang Yi, Chen Hu, Xia Wenjun, Chen Yang, Liu Baodong, Liu Yan, Sun Huaiqiang,
  and Zhou Jiliu.
\newblock {LEARN}++: Recurrent dual-domain reconstruction network for
  compressed sensing {CT}.
\newblock \emph{IEEE TRPMS}, 7:\penalty0 132--142, 2023.

\bibitem[Yu-Jung et~al.(2018)Yu-Jung, Alexandre, J., W., Sangtae, F., Simon,
  and Kris]{ctbfgs}
Tsai Yu-Jung, Bousse Alexandre, Ehrhardt~Matthias J., Stearns~Charles W., Ahn
  Sangtae, Hutton~Brian F., Arridge Simon, and Thielemans Kris.
\newblock Fast quasi-newton algorithms for penalized reconstruction in emission
  tomography and further improvements via preconditioning.
\newblock \emph{IEEE TMI}, 37:\penalty0 1000--1010, 2018.

\bibitem[Zang et~al.(2021)Zang, Idoughi, Li, Wonka, and Heidrich]{IntraTomo}
Guangming Zang, Ramzi Idoughi, Rui Li, Peter Wonka, and Wolfgang Heidrich.
\newblock Intratomo: Self-supervised learning-based tomography via sinogram
  synthesis and prediction.
\newblock In \emph{ICCV}, pages 1940--1950, 2021.

\bibitem[Zhang et~al.(2018)Zhang, Liang, Dong, Xie, and Cao]{ddnet}
Zhicheng Zhang, Xiaokun Liang, Xu Dong, Yaoqin Xie, and Guohua Cao.
\newblock A sparse-view {CT} reconstruction method based on combination of
  {DenseNet} and deconvolution.
\newblock \emph{IEEE TMI}, 37:\penalty0 1407--1417, 2018.

\end{thebibliography}
